# Personality as Relational Infrastructure:
# User Perceptions of Personality-Trait-Infused LLM Messaging

Dominik P. Hofer[1, 2, 3], David Haag[1], Rania Islambouli[1], Jan D. Smeddinck[1, 3]

1. Ludwig Boltzmann Institute for Digital Health and Prevention, Salzburg, Austria

2. Human Motion Analytics, Salzburg Research Forschungsgesellschaft, Salzburg, Austria

3. LMU Munich, Munich, Germany

Digital behaviour change systems increasingly rely on repeated, system-initiated messages to support users in everyday contexts. Large Language Models (LLMs) enable these messages to be personalised consistently across interactions, yet it remains unclear whether such personalisation improves individual messages or instead shapes users' perceptions through patterns of exposure. We explore this question in the context of LLM-generated Just-In-Time Adaptive Interventions (JITAIs), which are short, context-aware messages delivered at moments deemed appropriate to support behaviour change, using physical activity as an application domain. In a controlled retrospective study, ninety participants evaluated messages generated using four LLM strategies, including baseline prompting, few-shot prompting, fine-tuned models, and retrieval augmented generation, each implemented with and without the integration of Big Five Personality Traits to produce personality-aligned communication across multiple scenarios. Using ordinal multilevel models with within-between decomposition, we distinguish trial-level effects, capturing whether personality information improves evaluations of individual messages, from person-level exposure effects, capturing whether participants receiving higher proportions of personality informed messages exhibit systematically different overall perceptions Results showed no trial-level associations, but in this retrospective assessment, participants who received higher proportions of BFPT-informed messages rated the messages as more personalised, appropriate, and professional, and reported less negative affect. Between-person differences accounted for 54–81% of the variance across outcomes, indicating perceptions reflect exposure patterns rather than message fluctuations. We use Communication Accommodation Theory (CAT) as a post hoc interpretive lens for this pattern. CAT posits that when communicators converge toward each other's linguistic style, rapport builds cumulatively through repeated interactions rather than isolated exchanges. This may help explain why between-person variation in exposure to personality-aligned messages was associated with stronger effects than within-person message variation. While retrospective assessment allowed controlled comparison of LLM strategies within a single study, in-situ longitudinal studies are needed to validate these findings in real-world contexts. These preliminary results suggest that personality-based personalisation in behaviour change systems may operate primarily through aggregate exposure rather than per-message optimisation, with implications for how adaptive systems are designed and evaluated in sustained human AI interaction.

CCS CONCEPTS • Insert your first CCS term here • Insert your second CCS term here • Insert your third CCS term here

**ACM Reference Format:**
First Author's Name, Initials, and Last Name, Second Author's Name, Initials, and Last Name, and Third Author's Name, Initials, and Last Name. 2018. The Title of the Paper: ACM Conference Proceedings Manuscript Submission Template: This is the subtitle of the paper, this document both explains and embodies the submission format for authors using Word. In Woodstock '18: ACM Symposium

# 1 INTRODUCTION

Personalisation in digital behaviour change systems is often conceptualised as message-level optimisation: adapting content, tone, or framing to individual characteristics to improve immediate receptivity [1], [2], [3]. Yet interventions that initially perform well frequently fail to sustain engagement: novelty wears off, content becomes predictable, and users disengage [4], [5], [6], [7]. Frameworks such as the Multiphase Optimisation Strategy (MOST) [8] and Optimisation of Behavioural and Biobehavioural Interventions (ORBIT) [9] emphasise that intervention components may contribute differently to immediate versus sustained effects, warranting intervention generation designs that can isolate these contributions.

This raises a fundamental question: to what extent does personalisation improve evaluations of individual messages in a one-shot manner, versus shaping users' perceptions through broader exposure patterns across repeated interactions? These mechanisms are not mutually exclusive, but their relative contributions remain empirically unresolved—particularly for AI-generated interventions where scalable personalisation is now feasible.

We investigate this question using Big Five Personality Traits (BFPT) to personalise LLM-generated Just-In-Time Adaptive Interventions (JITAIs) for Physical Activity (PA). BFPT provide stable, interpretable dimensions that map to distinct communication preferences: openness correlates with exploratory framing, conscientiousness with structured messaging, and agreeableness with a warmer tone [10], [11], [12], [13]. By adapting language to users' trait profiles, systems may produce communication that feels familiar, potentially fostering rapport through consistent linguistic convergence rather than through optimising discrete interactions, which has been shown to be ineffective. [14].

We address three research questions:

**RQ1 (Trial-level, within-person):** Does incorporating BFPT information into LLM message generation improve user evaluations of individual JITAIs?

**RQ2 (Person-level):** Is higher aggregate exposure to BFPT-informed messages associated with more favourable overall perceptions of the system?

**RQ3 (Implementation):** Do different LLM generation strategies (baseline, few-shot, fine-tuned, RAG) differ in their associated user evaluations?

We conducted a controlled online study in which 90 participants retrospectively evaluated LLM-generated JITAIs across five personally relevant physical activity scenarios. Messages were generated using four approaches and implemented with and without BFPT integration. To mimic real-world implementation, LLMs independently decided whether to send a JITAI based on contextual information from each scenario, with each decision treated as independent and without access to prior messaging frequency. Using Bayesian ordinal multilevel models [15] with within-between decomposition, we separated trial-level effects, capturing whether BFPT integration improved evaluations of individual messages, from person-level effects, capturing whether between-person variation in the proportion of BFPT-informed messages was associated with overall perceptions.

Results revealed a clear dissociation between these levels of analysis: Trial-level associations between BFPT integration and message evaluations were null across all outcomes (all 97% HDIs crossed zero), suggesting personality integration did not detectably improve individual messages; However, person-level associations were substantial: Participants who received higher proportions of BFPT-informed messages rated them as more personalised, appropriate, and professional, and reported lower negative affect. Intraclass correlations showed that 54–81% of the variance was between-person, suggesting that perceptions were dominated by stable individual baselines. The person-level analysis further showed that these baselines varied systematically with aggregate exposure to BFPT-informed messaging.

To interpret the pattern of null trial-level but substantial person-level associations, we draw on Communication Accommodation Theory (CAT) [16] as a post-hoc interpretive framework. CAT states that accommodation effects emerge through sustained exposure rather than isolated exchanges, and that linguistic convergence fosters message acceptance. Recent extensions position CAT as applicable to human-AI interaction, suggesting that AI systems expressing consistent, personality-aligned communication may build rapport through analogous accommodation processes. While CAT did not inform our study design, it offers one potential explanation for our findings and a bridge to motivate further interdisciplinary research.

Our contributions are: (1) We provide empirical evidence that personality-based personalisation in LLM-generated JITAIs is not associated with differences at the level of individual messages, but is associated with *systematic differences in users' overall perceptions* as a function of exposure to personality-aligned messages; (2) We empirically compared multiple LLM generation strategies for personality-informed physical activity interventions, finding no consistent advantage of more resource-intensive approaches over simpler prompting methods; (3) We introduce *relational infrastructure* as a working concept to describe stable user features that scaffold consistent system accommodation across interactions, broadening design focus from per-message personalization to include the consistency and durability of accommodation stance.

## 2   RELATED WORK

Understanding how personalisation shapes human-AI interaction requires distinguishing between immediate effects (individual message optimisation) and cumulative effects (relationship development through sustained exposure). This distinction has received limited empirical attention despite the growing deployment of AI systems in contexts requiring sustained user engagement. We review literature establishing foundations for this work: (1) human-AI interaction and adaptive systems, establishing the design space for personalization; (2) Big Five Personality Traits as stable features with the potential to improve consistent personalization; (3) digital health interventions as our application domain; and (4) Communication Accommodation Theory as a viable post-hoc interpretive framework for understanding the person-level patterns that emerged in our data.

### 2.1   Human-AI Interaction and Adaptive Systems

As AI systems engage users in sustained interactions, research shows that relationship-like dynamics emerge [17]. Yet, most personalisation research assumes a transactional mechanism: better messages enhance immediate receptivity. Bickmore's Relational Agents Framework [18] offers an alternative in which effectiveness depends on whether personalisation demonstrates that the system "knows" the user through consistent relationship maintenance. Skjuve et al. [19] found chatbot interactions can foster companionship-like relationships, and Brandtzaeg et al. [20] demonstrated how users describe human-AI relationships, emphasising consistency and understanding. According to Streicher & Smeddinck, adaptive systems involve multiple design dimensions with variable levels, including *saliency* (whether adaptations are perceptible), *automation* (system- versus user-determined), and *explicitness* (implicit versus explicit control signals) [21]. The framework shows how complex and nuanced the design space of adaptive systems (e.g. for personalisation in applications for health) is and thereby implies the potential value in exploring how AIs might help navigate this design space in an ad-hoc, contextually-driven manner, as exhaustingly optimizing with in that complex space based on deterministic/heuristic systems has strong limits in scalability.

Furthermore, empirical evidence for immediate personalisation effects remains inconclusive, with some experimental trials showing small or null effects [22], [23]. This "personalisation paradox" [24] may reflect a predominant focus on immediate rather than cumulative effects, motivating our analytical approach and separating trial-level from person-level associations.

### 2.2 Big Five Personality Traits as Stable Personalisation Features

For personalisation to signal consistency, which is an important relational foundation, e.g. for trust [25], features must exhibit temporal stability, e.g. in terms of conversational style and tone. It can be hypothesised that BFPT can provide an operationalizable foundation for such stability [10], with personality-traits mapping to distinct traits that are mirrored/leave traces in communication preferences, e.g. with openness mapping to exploratory framing, agreeableness to warmer tone, and conscientiousness to structured messaging [11], [12], [13]. LLM research demonstrates the feasibility of personality expression, with generated text showing representative linguistic patterns [26], [11].

BFPT-informed generation creates an interesting configuration: traits are assessed explicitly (via a questionnaire) but integrated implicitly (users see only the resultant text). This low-saliency adaptation may accumulate into perceptible qualities over time [27]. However, single-message BFPT personalisation produces mixed results [28], [22] with modest reported trial-level effects but potential cumulative benefits, motivating the examination of both temporal levels.

### 2.3 Digital Health Interventions as Application Domain

We examine these dynamics in JITAIs for PA [3]. JITAIs involve repeated system-initiated contact, targeting behaviour change where trust matters, and increasingly leverage LLMs for message generation [29], [30]. Within the MOST [8] and ORBIT [9] frameworks, personality-informed generation represents a component whose contribution through immediate versus cumulative mechanisms can be isolated.

Critically, Park et al. [7], among other studies [1], [6], [31], demonstrated intervention fatigue as a cumulative between-person phenomenon, highlighting the importance as both a potential positivey – as well as negatively – contributing factor of temporal dynamics beyond immediate effects with regards to supporting behaviour change goals.

### 2.4 LLM Generation Approaches for Health Interventions

Different approaches exist for generating personalised health messages with LLMs. Wang et al. [32] compared multiple generation strategies for sleep health interventions, including baseline prompting, few-shot learning, and fine-tuning, and found that fine-tuned models produced higher-quality personalised recommendations. However, this contrasts with broader NLP research demonstrating that well-crafted prompts can achieve performance comparable to elaborate fine-tuning for many language tasks [33], [34], [35]. This tension around whether sophisticated, resource-intensive approaches outperform simpler methods for health personalisation with LLMs remains unresolved and has practical implications for deployment in resource-constrained settings. We therefore compare four generation strategies (baseline, few-shot chain-of-thought, fine-tuned, RAG) to contribute evidence on this question.

## 2.5 Communication Accommodation Theory as Relational Convergence Framework

We adopt CAT as a post-hoc framework for interpreting our results. CAT was not used to generate hypotheses; instead, the observed pattern of null trial-level associations alongside substantial person-level associations (cf. Results) led us to seek frameworks that account for cumulative/exposure proportion personalisation effects.

CAT posits speakers adjust communication style to achieve social goals, with convergence triggering similarity-attraction mechanisms through cumulative trust-building rather than single encounters [36]. Seventh-stage extensions include AI systems [16], with empirical work showing that LLMs exhibit convergence comparable to human dialogue [37], [38]. Critically, the "Adaptation Paradox" revealed that linguistic matching without a coherent persona underperforms static baselines [14], suggesting CAT convergence requires stable reference personality trait information, which BFPTs may provide.

To our knowledge, no prior work has simultaneously (a) manipulated BFPT infusion into LLM-generated interventions, (b) compared generation strategies with varying personality integration, and (c) employed multilevel designs separating trial-level from person-level effects. This gap matters for understanding how personalisation shapes human-AI interaction. If personality-based adaptation in conversational interaction primarily operates through cumulative exposure rather than through immediate improvements to messages, this would have substantial implications for how we design and evaluate adaptive systems. Our study addresses this gap through a retrospective, repeated-measures design where participants evaluate LLM-generated physical activity messages across multiple situations, enabling separation of immediate message effects from dosage exposure associations.

## 3 METHODOLOGY

We conducted two sequential studies: (1) a training data collection study that gathered human feedback on JITAIs across diverse scenarios and personality profiles, and (2) a retrospective assessment study that compared LLM-generated JITAIs with and without BFPT personalisation across multiple generation strategies. Figure *1* summarises the overall methodological procedure.

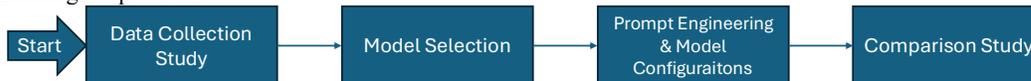

Figure 1: Summary of the methodological procedure

### 3.1 Training Data Collection

We recruited 210 participants via Prolific[1] (eligibility criteria: aged 18 or older, English-proficiency, and a self-reported sedentary lifestyle). Each participant evaluated five randomly selected scenarios from a pool of 15, as per Haag et al. [39]. Participants were asked to imagine themselves in hypothetical situations with embedded contextual details (time of day, activity, mood, physical state) and to rate how they would perceive corresponding JITAIs on 5 pt. Likert scales. In addition, participants provided qualitative feedback describing perceived strengths, weaknesses, and suggested modifications. Personality traits were assessed using the Big Five Inventory-10 (BFI-10) [40].Participant demographics were as follows: Age M=34.13, SD=13.58; 122 female, 88 male; ethnicity: 182 white, 9 asian, 8 black, 6 mixed, 5 Other. Personality trait distributions varied with no clear pattern (see Appendix 1 for complete BFPT profiles).

---

[1] https://www.prolific.com/

This mixed quantitative and qualitative dataset supported subsequent LLM fine-tuning, few-shot learning, and retrieval-augmented generation approaches. Dataset available at [BLIND_FOR_REVIEW]. Participants each received £9 compensation. The study was approved by the IRB of [BLIND_FOR_REVIEW].

### 3.2 Main Study: Retrospective JITAI Assessment

We employed a within-subjects retrospective design in which participants recalled five distinct time points from the past week based on a predefined contextual information collection text with blank elements to fill in and subsequently assessed LLM-generated JITAIs they hypothetically would have received in those situations.

For each recalled time point, participants completed the following steps. First, they specified the day and time, with no repetitions within participants. Second, they completed a contextual assessment covering 20 features, including current activity, location, body position, mood, stress, motivation, perceived barriers, weather, health status, recent and planned physical activity, and social context (see Appendix 2). Third, participants were presented with eight JITAI messages generated by different LLM configurations per scenario, described in Section 3.4, shown in randomised order using a tab-based interface. Finally, they rated each JITAI across multiple evaluative dimensions, described in Section 3.5. For each scenario, each LLM configuration first determined whether context was appropriate for sending a JITAI (receptivity decision), then generated message content only if sending was deemed appropriate. Each LLM decision was made independently using only the current turn context, without historical information. We thus evaluated both decision-making (when to send) and content generation (what to send, personalised or not).

Across 90 participants × 5 scenarios × 8 LLM configurations, we collected 3,600 send/no-send decisions (2,345 send) and 77,280 item ratings (21 items per sent JITAI). Figure 2 illustrates the study procedure.

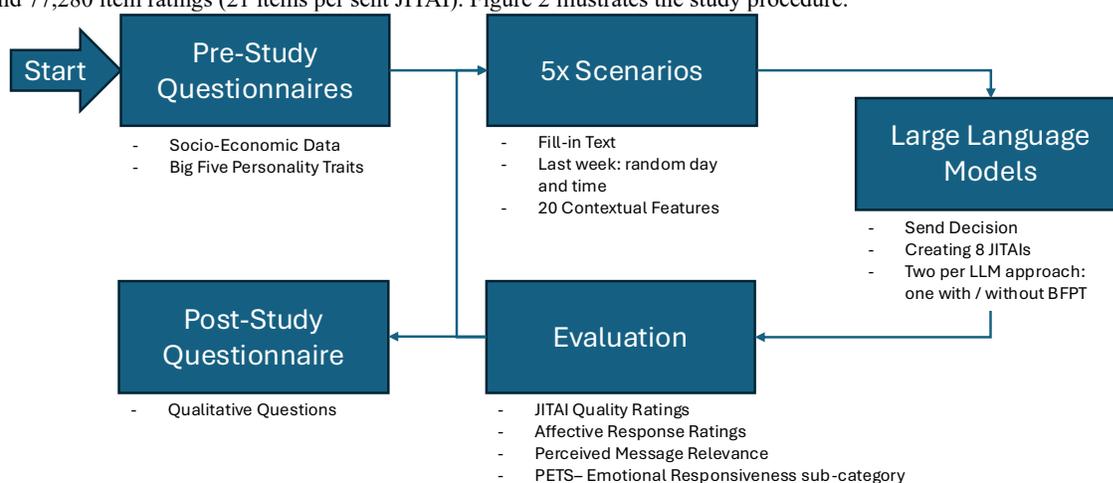

Figure 2: Procedure of Comparison Study

### 3.3 Participants

We recruited two participant cohorts via Prolific, using eligibility criteria identical to those of the training study, with the additional requirements of computer access (no mobile devices) and a stable internet connection due to technical requirements.

**Pre-screening:** To ensure receptivity relevance, we administered a screening question: "I intend to be physically active during my leisure time over the next 5 weeks" (yes/no). Only participants answering "yes" were invited to the main study.

**Final sample:** 90 participants completed the study successfully. Age: M=37.20, SD=10.5, range 22-72; 45 female, 45 male; ethnicity: 80 White, 5 Asian, 3 Black, 2 Mixed. Personality traits showed varied distributions across all Big Five dimensions (complete profiles in Appendix 1).

Participants received compensation at Prolific standard rates. The study was approved by the IRB of [BLIND_FOR_REVIEW].

### 3.4 LLM Configuration

We selected ChatGPT-4o (chatgpt-4o-latest, OpenAI, July 2025), based on benchmark performance in JITAI generation tasks and emotional intelligence evaluations [41]. While OpenAI's o3 occasionally outperforms ChatGPT-4, substantially longer response times made it impractical for near real-time message generation. Four generation strategies were compared, each implemented both with and without BFPT information (8 configurations total):

- **Baseline:** Minimal prompt engineering. Model instructed to act as "*AI healthcare expert for motivational messages*" and incorporate any provided BFPT information.
- **Few-Shot Chain-of-Thought with Self-Consistency (FS-CoT-SC):** Following Kim et al. [42] the model receives three randomly selected user feedback examples from the training data, generates multiple parallel reasoning chains to evaluate what to send, and aggregates them into a coherent response.
- **Retrieval-Augmented Generation (RAG):** For each query, the system retrieves the three most contextually similar examples from the training dataset using Universal Sentence Encoder v3 embeddings. BFPT is prioritised when available; otherwise, general contextual similarity is used. Retrieved examples guide response generation.
- **Fine-Tuning:** Models trained on the training dataset using the OpenAI supervised fine-tuning workflow. Two versions created: with and without BFPT training data. Prompt structure mirrored RAG but relied on learned weights rather than explicit retrieval.

Complete prompts for all configurations appear in Appendix 3. Data for prompt generation and model training are described in Section 3.1. The technical implementation used Python with Streamlit (UI) and the OpenAI API (backend). Code available at [BLIND_FOR_REVIEW].

### 3.5 Measures

We implemented a multi-dimensional evaluation combining validated and exploratory instruments, which were applied for every sent JITAI:

- **Primary Outcomes - Perceived Personalisation:**
  - Perceived Message Relevance Scale [43]: Four items assessing (1) personal relevance, (2) contextual fit, (3) generic content (reverse-coded), and (4) lack of customisation (reverse-coded). Labelled p1-p4 in results.
  - PETS Emotional Responsiveness subscale [44] Six items measuring perceived socio-emotional sensitivity on a 0-100 pt quasi-continuous scale.
- **Secondary Outcomes - JITAI Quality:** Adapted from Haag et al. [39]: appropriateness (sent), appropriateness (content), engagement, effectiveness, and professionalism. See Appendix 4 for question per item.
- **Secondary Outcomes - Affective Response:** Emotional intensity ratings for seven emotions, also from Haag et al. [39]: anger, annoyance, frustration, happiness, sadness, fear, surprise.

All items used 7-point Likert scales (1=not at all, 7=extremely) except PETS (0-100 continuous).

## 3.6 Statistical Analysis

Our analytical approach comprised three stages to test whether BFPT personalisation operates through trial-level message optimisation or person-level dosage exposure. First, we characterised the distributions of outcomes to guide modelling choices. Second, we identified and controlled confounding variables arising from non-random BFPT exposure. Third, we decomposed personalisation effects into trial-level (immediate) and person-level (exposure proportions) components using Bayesian multilevel ordinal regression [45].

### 3.6.1 Descriptive Analysis and Normality Assessment

We characterised the distributions of the outcome variable using descriptive statistics and histograms. Mainly, non-normal distributions guided the selection of ordinal regression over parametric approaches.

### 3.6.2 Confounding Control

Because LLMs used contextual information for send decisions, BFPT exposure was non-random, creating potential confounding. We addressed this through multicollinearity assessment:

- **Feature preparation:** Continuous variables (Big Five traits, psychological states) were z-standardised. Highly imbalanced categorical variables (e.g., health status: 82% "neither sick nor on holiday") were binarised. Free-text features (e.g., "what activity are you doing?") were categorised following validated frameworks (e.g., Rapid Assessment of Physical Activity Questionnaire [46] for activity intensity). Example categorisation schemes appear in Appendix 5.
- **Variance Inflation Factor (VIF) analysis:** We iteratively refined predictors to: (1) retain substantively important features (health status, PA scheduling, activity type, Big Five traits), (2) minimise multicollinearity (target VIF < 10), and (3) maintain parsimony. The resulting final predictor set includes: *user health status* (binary), *participant type* (new vs. returning), *planned PA tomorrow* (4 categories), *current activity* (4 categories), *PA performed today* (4 categories), *Big Five traits* (z-scored). Most predictors achieved VIF < 2; current activity reached VIF = 8-36 across categories, but did not inflate uncertainty in focal predictors (trial-level and person-level BFPT effects both VIF < 2).
- **Verification of BFPT independence:** We examined correlations between participant-level BFPT proportion (average exposure across scenarios; Appendix 6) and all contextual/personality predictors. Correlations were uniformly weak (all $r < 0.15$; Appendix 6), and the BFPT proportion varied minimally across contextual strata (range: 0.47-0.49). While LLMs used context to inform timing, they didn't systematically favour BFPT messages for specific participant profiles, thereby minimising concerns of confounding bias.

### 3.6.3 Bayesian Multilevel Ordinal Regression

We evaluated both linear and ordinal regression approaches. Linear models showed poor convergence ($\hat{R} > 1.01$, ESS < 1,000) and excessive coefficient variability (CV > 100%) due to bounded ordinal data with non-normal distributions. Ordinal regression is statistically appropriate for Likert scales [15], [47], [48], [49] and achieved excellent convergence (all $\hat{R} = 1.0$, ESS > 1,000); we therefore report only ordinal results.

We implemented Bayesian multilevel ordinal regression using Bambi (Bayesian Model-Building Interface in Python) [45], which handles hierarchical structures with ordinal outcomes and non-parametric distributions. The model formula is: `outcome ~ [confounders] + bfpt_person_mean + trial_level + (1 + trial_level|participant_id)`.

Table 1: Description of formula components for ordinal regression

| Component | Component Description |
|---|---|
| outcome | User answers of the primary and secondary outcome variables per JITAI, e.g. p1-p4, appropriateness content, and happy |
| confounders | Confounding features, e.g. Big Five traits (z-scored), health status, participant type, PA scheduling, current activity |
| bfpt_person_mean | Each participant's individual proportion/dosage of BFPT-informed (vs non-BFPT-informed) messages across all scenarios (person-level exposure proportion) |
| trial_level | Within-person deviation from each participant's mean BFPT exposure (trial-level personalisation effect) |
| random effects | Participant-specific intercepts (baseline perception) and slopes for trial_level (individual sensitivity to trial-level BFPT variation), (1 + trial_level | participant_id) |

This within-between decomposition tests two distinct hypotheses:

1. **Trial-level associations:** Whether receiving a BFPT-personalised message shows higher perceived ratings of that specific message relative to non-BFPT-personalised messages for the same individual (*immediate message optimisation*).
2. **Person-level associations:** Whether individuals who received higher proportions/dosage of BFPT-informed messages show systematically different overall ratings (*person-level exposure proportion*).

Person-level BFPT was not standardised despite narrow distribution (40-57%, SD=0.03) because standardisation impaired MCMC convergence [50] (ESS decreased 40%, R̂=1.01). Unstandardized coefficients retained interpretable magnitudes (40 – 57% of bfpt infused message dosage) and yielded stable sampling [50] – cf. Results section. Empathy outcomes used a beta transformation [51] due to a bounded 0-100 scale and severe skew. We assessed convergence using R̂ statistics (target < 1.01) and effective sample sizes (target > 1,000) [50]. We calculated intraclass correlation coefficients (ICC) from posterior draws to quantify variance partitioning between-person vs. within-person components.

The narrow person-level BFPT range limits our ability to detect nonlinear effects or to establish optimal exposure thresholds; this constraint arises from the naturalistic LLM decision-making design (autonomous send/no-send decisions) rather than from experimental manipulation of exposure levels.

All analyses used 97% highest density intervals (HDI) for credible effects. Effects were considered credible if HDIs excluded zero [48].

## 4 RESULTS

We collected 75,600 data points from 90 participants (21 items per JITAI × 8 JITAIs per scenario × 5 scenarios per participant). Across conditions, the eight LLM configurations made 3,600 send/no-send decisions, resulting in 2,345 delivered JITAIs and 1,255 withheld messages. Contextual factors in scenarios that were not always applicable (e.g. PA performed today, and PA planned tomorrow) were coded as "Other" when irrelevant to the specific scenario context. We defined empathy and p1–p4 as primary outcomes. Secondary outcomes included JITAI Quality measures (appropriateness, engagement, effectiveness, professionalism) and Affective Response scales.

### 4.1 Descriptive:

JITAI quality ratings ranged from 4.72 to 5.26 on 7-point scales, with moderate variability (SD = 1.63–1.83). Negative affective responses (anger, annoyance, frustration, sadness, fear) were rare (M = 1.25–1.84) and heavily right-skewed, while happiness was moderate (M = 4.05, SD = 1.99) with a more even distribution. Empathy varied considerably between individuals (M = 60.09, SD = 24.74, range = 0–100), with a right-skewed distribution. Personalisation ratings displayed opposite skewness: left-skew for p1 and p2 (M = 4.89–4.97, SD = 1.78–1.80), right-skew for p3 and p4 (M = 2.40–2.52, SD = 1.67–1.76). A tabular and figure representation can be found in Appendix 7.

### 4.2 Addressing Confounding Through Multicollinearity Analysis

Because LLM send-decisions were informed by contextual information, exposure to BFPT-informed messages was not randomly assigned, raising the possibility of confounding. To assess whether participant-level BFPT exposure was systematically associated with contextual or individual characteristics, we examined multicollinearity among all candidate predictors.

Variance inflation analyses indicated no multicollinearity for focal predictors. In particular, participant-level BFPT exposure showed only weak associations with contextual and personality covariates (all $|r| < .15$). While some contextual variables exhibited higher collinearity, this did not inflate uncertainty in the trial-level or person-level BFPT estimates. Together, these diagnostics suggest that observed person-level associations with BFPT exposure are unlikely to be explained by confounding due to contextual or individual differences. Full diagnostic details are reported in Appendix 8.

### 4.3 LLM Generation Strategy Effects

We examined whether generation strategy affected outcomes through ordinal regression, focusing on model type and interactions between model type and person-level BFPT exposure proportion (all with confounding controls). Main effects showed no credible differences between strategies (all 97% HDIs included zero; Appendix 13).

For interaction effects, while several model type × bfpt_person_mean interactions reached statistical credibility (Appendix 13), no consistent interpretable pattern emerged: effects for conceptually inverse outcomes (e.g., Relevant vs. Doesn't Fit) were not symmetric across strategies, and interactions appeared across both primary and secondary outcomes without theoretical coherence. Given the number of comparisons tested and the absence of alpha correction for repeated measures, we interpret these as statistical variability rather than robust differential effects.

### 4.4 Bayesian Multilevel Ordinal Regression

To distinguish whether personality-based personalisation operates at the level of individual messages or through dosage exposure across interactions, we fit Bayesian multilevel ordinal regression models with explicit within–between decomposition. The results can be found in Figures 4 & Figure 5, as well as in Table 6 and Table 7 (in Appendix 10). All models achieved excellent convergence ($\hat{R} = 1.0$) and adequate effective sample sizes (ESS > 1,000). No outcome variable demonstrated credible trial-level associations—all 97% HDIs included zero. Effect sizes were uniformly small ($|\beta| < 0.07$), suggesting no detectable association between *BFPT personalisation in individual messages* and *immediate ratings*. In contrast, person-level associations were observed for 14 of 17 outcomes:

**Perceived personalisation:** p1 ($\beta = 3.56$, HDI [0.77, 6.46]) and p2 ($\beta = 3.73$, HDI [0.73, 6.50]) showed positive associations. In contrast, p3 ($\beta = -8.28$, HDI [-11.96, -4.42]) and p4 ($\beta = -4.11$, HDI [-7.04, -1.20]) showed negative associations—indicating positive associations between BFPT exposure and perceptions of personalisation, and negative associations with perceptions of generic content.

**JITAI quality dimensions:** Appropriateness showed robust associations (sent: $\beta = 4.63$, HDI [1.89, 7.43]; content: $\beta = 4.90$, HDI [2.17, 7.79]), as did professionalism ($\beta = 5.66$, HDI [2.76, 8.59]), effectiveness ($\beta = 4.09$, HDI [1.36, 7.03]), and engagement ($\beta = 3.24$, HDI [0.46, 6.07]).

**Negative affect:** Strongest person-level associations were observed for scared ($\beta = -16.22$, HDI [-25.56, -7.29]), sad ($\beta = -10.68$, HDI [-15.97, -5.70]), angry ($\beta = -8.02$, HDI [-12.11, -4.03]), frustrated ($\beta = -6.62$, HDI [-10.21, -3.20]), and annoyed (β = -4.26, HDI [-8.39, -0.37])—all showing negative associations between BFPT exposure and negative emotional responses.

**Non-significant outcomes:** Empathy (β = 0.849, HDI [-5.29, 7.25]), happiness (β = 2.99, HDI [-0.34, 5.79]), and surprise (β = -2.17, HDI [-4.98, 0.74]) showed person-level associations that were not credibly different from zero.

The pattern of null trial-level associations alongside substantial person-level associations is consistent with CAT predictions. However, the cross-sectional design cannot distinguish cumulative trust-building from stable individual differences in personality-content receptivity.

### 4.5  Intraclass Correlation: Between-Person Stability

Across outcomes, intraclass correlation coefficients indicated that the majority of variance in user ratings was attributable to stable between-participant differences rather than message-level variation. ICCs ranged from 0.54% to 0.82%, demonstrating substantial between-person stability across perceived personalisation, JITAI quality, and affective responses. In more detail, *scared* showed the highest ICC (0.82, 97% CI [0.73, 0.89]). Other affective responses demonstrated substantial stability: sadness (0.67, CI [0.58, 0.77]), anger (0.63, CI [0.53, 0.73]), surprise (0.61, CI [0.53, 0.70]). JITAI quality showed moderate-to-high ICCs: effectiveness (0.61, CI [0.53, 0.69]), professionalism (0.59, CI [0.51, 0.67]), engagement (0.57, CI [0.48, 0.65]), and appropriateness (0.55–0.56). Personalisation dimensions (personal – doesn't fit).

Empathy showed a notably lower ICC (0.23 vs. 0.54–0.82 for other outcomes), indicating greater within-person variability. This may reflect the PETS subscale's continuous 0–100 format, which allows finer discrimination, or it may reflect the fact that perceived empathy responds more to momentary message features than to stable person-level factors. This warrants further investigation, as empathy may operate through mechanisms distinct from those of other socio-emotional perceptions. Complete the table and barplot in Appendix 11.

### 4.6  Dose-Response Plots

To translate regression results into interpretable effects, we generated posterior predictions across the observed range of BFPT exposure proportions (40–57%) and examined the distributions of the observed data for the lowest (Q1: 40–44%) and highest (Q4: 52.5–57%) exposure quartiles. Figure 7 presents dose-response curves (model predictions) and observed quartile means. We separated positive outcomes (higher ratings desired) from negative outcomes (lower ratings desired) for clarity.

**For positive outcomes,** the posterior predictions show consistent improvement from Q1 to Q4. Most quality dimensions started at 4.6–5.0 (Q1) and increased to 4.8–5.5 (Q4), representing approximately a 0.4-point improvement. Happiness showed a similar trajectory (3.9 → 4.3, Δ ≈ +0.4). Observed quartile means confirmed this pattern: all Q4 means exceeded Q1 means by 5–10%.

**For the negative outcomes,** the posterior predictions showed weaker but consistent decreases. "Generic" and "doesn't fit" started at ~2.5–2.75 and decreased to ~2.0–2.25. Annoyed, frustrated, and angry started at ~2.0 (±0.1) and decreased to ~1.6 (±0.1). Sad and scared started at ~1.6 and decreased to ~1.2. Surprise showed minimal change, consistent with non-significant person-level associations.

A significant interpretive challenge arose for negative affect: observed Q4 means sometimes surpassed Q1 means, contradicting model-predicted decreases. This apparent reversal results from severe floor effects—over 70% of participants rated negative emotions at 1 regardless of condition, leaving little observable variation. The regression coefficients reflect effects among the minority who display heightened negative affect, but they do not support overall reductions at the population level based on the raw data. Therefore, we interpret the negative affect findings with caution: the statistical pattern indicates different sensitivity to BFPT exposure among those predisposed to negative reactions, but claims of broad negative affect reduction should be viewed cautiously, as unwarranted.

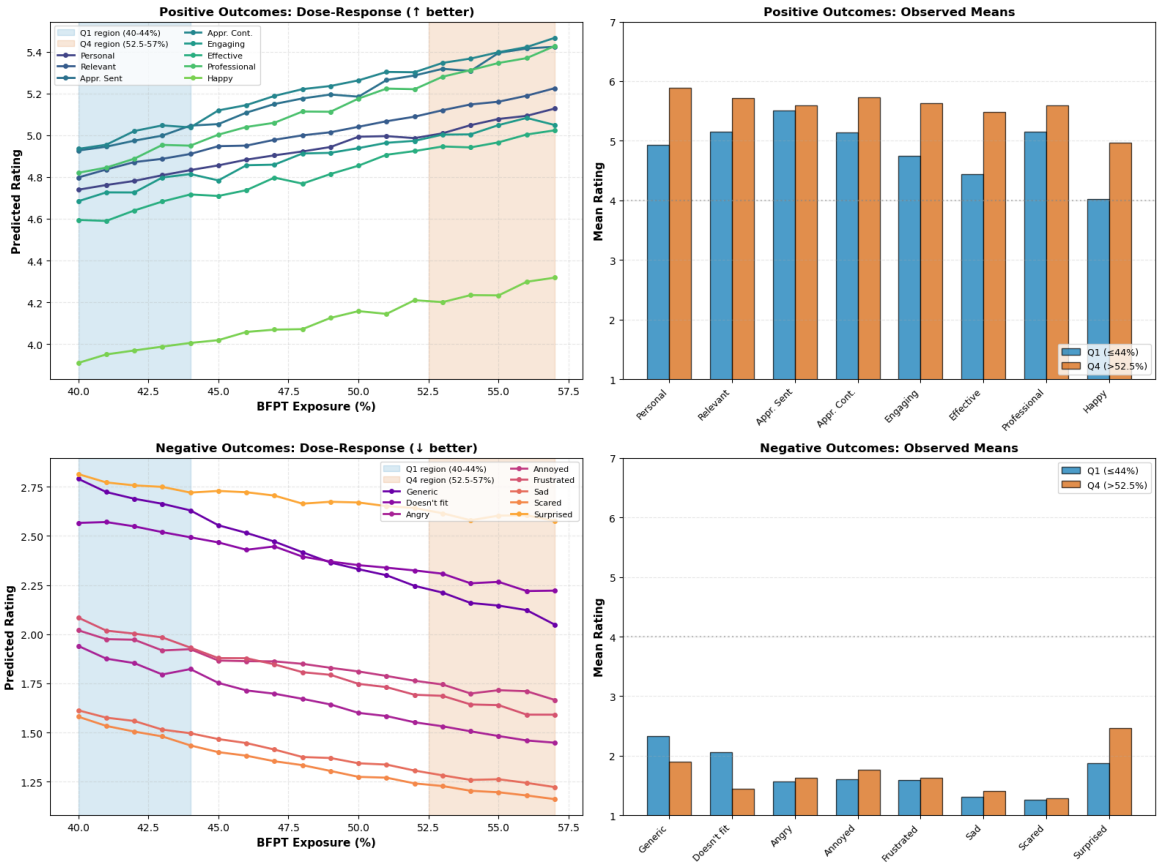

Figure 3: Dose-Response (left) and Observed Means (right) of the Outcome Variables

This suggests that personality-congruent messaging primarily enhances positive perceptions rather than reducing already low negative affect. For a full listing of observed Q1 vs. Q4 distributions cf. to Appendix 12.

Results demonstrate null trial-level associations alongside strong person-level associations for 14 of 17 outcomes, with dose-response patterns showing progressive improvements from 40% to 57% BFPT exposure. High ICCs (54–82%) confirm these reflect stable between-person perceptions rather than trial-level fluctuations. Negative emotion floor effects limit the observable magnitude of absolute changes despite credible statistical effects.

## 5 DISCUSSION

This study examined whether personality-based linguistic adaptation in human-AI interaction operates through immediate message optimisation and/or exposure proportions. Our findings suggest the latter: person-level associations emerged while trial-level associations were null, and different LLM generation strategies performed equivalently. Below, we interpret these patterns, discuss potential implications for adaptive system design, and outline directions for future research.

## 5.1 Interpretation: Personalisation as Relational Process

Participants who received higher proportions of personality-aligned messages showed more favourable perceptions. The disconnect between person-level and trial-level associations suggests that including BFPT information is associated with a more favourable perception of JITIs with increased dosage, rather than a one-shot improvement of message perception.

ICCs revealed 54-81% of variance was between-person for 16 of 17 outcomes (median ICC = 0.59). This indicates perceived personalisation, appropriateness, and engagement function as stable perceptions shaped by higher dosage rather than momentary message features. Overall, user perception profiles seem to have remained stable across messages.

The pattern across exposure quartiles, that participants in higher BFPT proportion quartiles rated messages more favourably, is consistent with our person-level association interpretation. However, our limited exposure range (40-57%) prevents us from determining whether effects plateau, continue, or reverse at higher levels.

Empathy showed notably lower ICC (0.23) and null person-level associations. This may reflect the PETS subscale's continuous format, enabling finer discrimination, or indicate that perceived empathy responds to different features than the other outcomes. This warrants further investigation.

Floor effects complicate the interpretation of negative affect findings. Over 70% of participants rated negative emotions at the lowest level regardless of exposure. Credible person-level associations emerged from a minority of participants who showed elevated negative responses, limiting generalizability.

## 5.2 Generation Strategy Equivalence

Resource-intensive LLM strategies (fine-tuning, RAG) did not outperform simple baseline prompting across main effects, and exploratory interaction analyses revealed no consistent patterns suggesting strategy-dependent BFPT associations.

One interpretation is that the consistency of BFPT integration matters more than the sophistication of generation. This aligns with findings that well-structured prompts can match fine-tuned model performance for many tasks [33], [34], [35], though it contrasts with Wang et al.'s [32] finding that fine-tuning improved sleep health recommendations.

However, the retrospective design may lack sensitivity to detect differences that emerge over extended use. Strategy differences might manifest in dimensions we did not assess, such as linguistic diversity, long-term engagement, or robustness across contexts. These preliminary observations suggest simpler approaches warrant investigation in longitudinal settings before drawing implementation conclusions.

## 5.3 Implications for Human-AI Interaction

The following implications are conditional on longitudinal validation; our retrospective design gives a formative nature to the work at hand and motivates such studies to possibly solidify design implications.

**Position in the adaptive systems design space:** The null trial-level effects combined with positive person-level associations suggest an intriguing configuration within the design space for adaptive systems [27]. Low-saliency, implicit personalisation may function differently than explicitly signalled adaptations: rather than producing immediate recognition that "this message was tailored for me," accumulated exposure to personality-congruent communication may shape global perceptions of system quality without users consciously attributing this to personalisation. This aligns with observations that adaptive systems must navigate tensions between user agency and automated optimisation [21], [27], [52]. Personality-informed generation may represent a form of background adaptation that supports user experience without demanding attention or explicit control.

**Temporal reframing of evaluation:** If our between-person findings reflect genuine cumulative exposure effects, evaluation frameworks may need to combine immediate response metrics with longitudinal indicators. The possibility that users do not consciously detect personality alignment in individual messages yet show favourable perceptions with higher exposure proportions motivates such longitudinal investigation. This would parallel human relationship development: trust/liking accumulates through consistent behaviour and getting to know a person over time, rather than isolated interactions. This, therefore, also motivates a comprehensive psycho-socially (and linguistically) informed angle on human-AI interaction that may be framed as "*Human-AI Relations*".

**Connections to therapeutic alliance:** The person-level patterns parallel findings from therapeutic alliance research, where relationship quality rather than specific techniques often predicts outcomes across interventions [53]. If sustained personality-congruent communication functions analogously to therapeutic responsiveness, AI-delivered interventions may benefit from design approaches that include relational consistency, aligning with calls to consider "digital therapeutic alliance" in technology-mediated support [54]. At the same time, it will be important to also critically consider possible negative (side-)effects of such "AI personal-conversation mimicry" both in terms of individual human-to-AI interaction effects (cf. the emerging concept of "AI Psychosis"), as well as potential tensions arising through increasingly agentically represented and conversationally interacting AI as a "third actor" challenging established patient-provider relationships.

## 5.4    Implications for Transparency and User Control

The trial- and person-level findings raise questions about the transparency and user control of AI systems. If cumulative exposure to personality-aligned communication shapes user perceptions without explicit awareness, what are the implications for informed consent? Current AI transparency paradigms often focus on explaining individual decisions [55]. Our findings suggest that *relational effects* accumulating across interactions may be equally essential yet harder to detect. Future work should explore whether making personality-based adaptation explicit enhances or undermines accommodation effects, and how users can be empowered to understand and control personality-based adaptive conversation in sustained AI interactions.

## 5.5    Communication Accommodation Theory as a post-hoc Interpretive Lens

CAT posits that linguistic convergence triggers similarity-attraction mechanisms through repeated demonstrations of understanding [36]. The stability of personality traits may enable this by providing coherent identity signalling: unlike turn-by-turn mimicry that can violate expectations [14], BFPT-based adaptation maintains a consistent persona while demonstrating user understanding.

We apply this framework cautiously. Our cross-sectional design cannot conclusively confirm the temporal mechanism implied by CAT, but the outcomes of the ordinal regression as well as the ICC show measured patterns that are consistent with its predictions. 'Person-level' in our analysis refers to between-person variation in BFPT message proportion within a single evaluation session, not within-person change over time. CAT's convergence mechanism implies the latter. Demonstrating actual cumulative trust-building requires longitudinal designs tracking individuals across genuine repeated interactions.

Alternative interpretations remain viable: global impression formation independent of accommodation, demand characteristics, or differential sensitivity, where some individuals are more receptive to personality-aligned content. Field studies with randomised exposure levels and process measures are needed to more firmly distinguish these explanations.

## 5.6    Relational Infrastructure

Motivated by these patterns, we introduce the term *relational infrastructure* to describe stable user features (e.g., Big Five traits) that can scaffold positively perceived accommodation signals across interactions that can be successfully operationalized by generative AI foundation models, particularly LLMs. Unlike optimisation parameters that are tuned to improve individual message quality, relational infrastructure provides the foundation for maintaining a coherent system identity and demonstrating ongoing user understanding and learning via continuous interactions. This framing broadens design focus from per-message personalisation to also more speficially consider aspects such as the consistency and durability of the accommodation stance.

### 5.7  Future Research / Next Steps

Multiple future research directions are implied by this work. First, longitudinal field studies using micro-randomized trials [4] could randomly assign BFPT-informed versus standard messages within individuals over time, enabling causal inference about cumulative exposure effects while preserving ecological validity. This design would disentangle whether personality alignment produces genuine within-person changes or merely reflects stable individual differences in message receptivity. Second, cross-cultural validation is needed, as prior work shows cultural factors moderate accommodation dynamics [37]. Communication norms and power distance may alter how personality-based adaptation operates across populations. Third, other temporally stable individual differences beyond BFPT warrant exploration. Constructs from Self-Determination Theory [56], Regulatory Focus [57], or Health Locus of Control [58] could provide complementary personalisation mechanisms for digital health interventions.

### 5.8  Limitations

Several limitations apply and should be considered when interpreting our findings. First, our retrospective single-session design limits causal inference in three ways: (1) BFPT exposure emerged from LLM decision-making rather than experimental manipulation, creating potential confounding by unmeasured individual differences that influenced both messaging decisions and participant ratings; (2) participants rated hypothetical messages rather than receiving real interventions during daily life; and (3) our 'person-level' associations reflect between-person differences at one timepoint, not within-person change over time. Establishing causal cumulative mechanisms requires longitudinal field studies with experimental manipulation of exposure density. Second, BFPT exposure varied narrowly (40-57%), limiting detection of nonlinear effects or identification of optimal personalisation thresholds. This constraint arose from naturalistic LLM message-sending decisions rather than from systematic variation in dosage across participants. Third, our predominantly white Prolific-recruited sample limits generalizability. Cross-cultural validation is essential, as personality expression and communication preferences likely vary across cultural contexts. Fourth, our operationalisation of BFPT 'dose' as proportional exposure (percentage of BFPT-personalised messages) does not account for absolute message counts, temporal spacing, or ordering effects. Participants receiving identical proportions may have experienced different absolute exposures, and the timing or sequence of BFPT messages may independently influence cumulative effects, independent of overall dosage. Fifth, our LLM strategy comparison was limited to specific architectural approaches (baseline, few-shot, fine-tuned, RAG). Findings may not generalise to other prompting strategies, model architectures, or personalisation frameworks. Finally, while our observed person-level effects are aligned with Communication Accommodation Theory's prediction that adaptation operates cumulatively, our design cannot confirm the temporal accumulation mechanism that CAT proposes. Our findings provide preliminary evidence warranting longitudinal validation, but cannot yet establish that personality-based personalisation builds relational bonds over time.

## 6 CONCLUSION

This study investigated whether personality-based linguistic adaptation of LLM conversational choices in behaviour change support works through immediate message optimisation, between-person differences in exposure to personality-aligned content, or both. The pattern of no trial-level associations alongside significant person-level ones suggests that participants who received more personality-aligned messages perceived the system more positively. We contextualise how this finding may become increasingly relevant as AI systems engage users over longer periods. The Big Five Personality Traits may be particularly suitable for this function, as their stability over time allows for consistent linguistic adaptation that correlates with positive perceptions of system quality, regardless of trial-level differences. We applied Communication Accommodation Theory post-hoc to interpret these results, especially the idea that linguistic convergence between communication partners increases message acceptance. Our study design did not facilitate natural convergence, and we encourage further research to explore whether genuine bidirectional adaptation over longer-term and multi-turn conversations between users and AI systems produces similar outcomes. Longitudinal field studies with randomised exposure conditions are vital before these initial findings can inform design practices. If confirmed, they suggest that assessing personality-based adaptation should focus on long-term relationship indicators rather than immediate response metrics, and that consistent accommodation may be more critical than per-message optimisation. As human-AI interactions become more sustained, understanding with more psycho-social and linguistically informed nuance, under a broad theoretical framing for which we propose the term "Human-AI Relations", how stable user traits support effective personalisation is increasingly critical. Accordingly, we propose the concept of relational infrastructure for personality-trait information that can be effectively operationalised by LLMs.

# A  APPENDICES

## A.1 Table of BFPT Distribution of Participants

Table 2: BFPT Distribution for datacollection study

| Numeric Value | Openness to Experience | Absolute / Relative Value | Conscientiousness | Absolute / Relative Value | Extraversion | Absolute / Relative Value | Agreeable | Absolute / Relative Value | Neurotic | Absolute / Relative Value |
|---|---|---|---|---|---|---|---|---|---|---|
| 5.0 | open | 26 / 12.4% | conscientious | 13 / 6.2% | extroverted | 4 / 1.9% | agreeable | 13 / 6.2% | neurotic | 34 / 16.2% |
| 4.5 | primarily open | 32 / 15.2% | primarily conscientious | 19 / 9.0% | primarily extroverted | 7 / 3.3% | primarily agreeable | 35 / 16.7% | primarily neurotic | 32 / 15.2% |
| 4.0 | slightly open | 27 / 12.9% | slightly conscientious | 36 / 17.1% | slightly extroverted | 14 / 6.7% | slightly agreeable | 40 / 19.0% | slightly neurotic | 24 / 11.4% |
| 3.5 | barely open | 35 / 16.7% | barely conscientious | 21 / 10.0% | barely extroverted | 17 / 8.1% | barely agreeable | 35 / 16.7% | barely neurotic | 26 / 12.4% |
| 3.0 | neither/nor | 52 / 24.8% | neither/nor | 52 / 24.8% | neither/nor | 28 / 13.3% | neither/nor | 34 / 16.2% | neither/nor | 28 / 13.3% |

| Numeric Value | | | | | | | | | | | | |
|---|---|---|---|---|---|---|---|---|---|---|---|---|
| 2.5 | barely closed | 18 / 8.6% | barely unconscientious | 48 / 22.9% | barely introverted | 43 / 20.5% | barely antagonistic | 25 / 11.9% | barely emotionally stable | 23 / 11.0% | | |
| 2.0 | slightly closed | 15 / 7.1% | slightly unconscientious | 19 / 9.0% | slightly introverted | 39 / 18.6% | slightly antagonistic | 19 / 9.0% | slightly emotionally stable | 25 / 11.9% | | |
| 1.5 | primarily closed | 5 / 2.4% | primarily unconscientious | 2 / 1.0% | primarily introverted | 38 / 18.1% | primarily antagonistic | 5 / 2.4% | primarily emotionally stable | 10 / 4.8% | | |
| 1.0 | closed | 0 / 0.0% | unconscientious | 0 / 0.0% | introverted | 20 / 9.5% | antagonistic | 4 / 1.9% | emotionally stable | 8 / 3.8% | | |

Table 3: BFPT distribution for the Comparison Study

| Numeric Value | Openness to Experience | Absolute / Relative Value | Consci-entiousness | Absolute / Relative Value | Extra-version | Absolute / Relative Value | Agreeable | Absolute / Relative Value | Neurotic | Absolute / Relative Value |
|---|---|---|---|---|---|---|---|---|---|---|
| 5.0 | open | 14 / 15.2% | conscientious | 13 / 14.1% | extroverted | 2 / 2.2% | agreeable | 8 / 8.7% | neurotic | 10 / 10.9% |
| 4.5 | primarily open | 11 / 12.0% | primarily conscientious | 13 / 14.1% | primarily extroverted | 3 / 3.3% | primarily agreeable | 12 / 13.0% | primarily neurotic | 5 / 5.4% |
| 4.0 | slightly open | 17 / 18.5% | slightly conscientious | 15 / 16.3% | slightly extroverted | 8 / 8.7% | slightly agreeable | 20 / 21.7% | slightly neurotic | 16 / 17.4% |
| 3.5 | barely open | 18 / 19.6% | barely conscientious | 16 / 17.4% | barely extroverted | 12 / 13.0% | barely agreeable | 24 / 26.1% | barely neurotic | 12 / 13.0% |
| 3.0 | neither/nor | 24 / 26.1% | neither/nor | 16 / 17.4% | neither/nor | 9 / 9.8% | neither/nor | 17 / 18.5% | neither/nor | 14 / 15.2% |
| 2.5 | barely closed | 3 / 3.3% | barely unconscientious | 11 / 12.0% | barely introverted | 19 / 20.7% | barely antagonistic | 6 / 6.5% | barely emotionally stable | 8 / 8.7% |
| 2.0 | slightly closed | 3 / 3.3% | slightly unconscientious | 5 / 5.4% | slightly introverted | 12 / 13.0% | slightly antagonistic | 3 / 3.3% | slightly emotionally stable | 14 / 15.02% |
| 1.5 | primarily closed | 2 / 2.2% | primarily unconscientious | 2 / 2.2% | primarily introverted | 18 / 19.6% | primarily antagonistic | 1 / 1.1% | primarily emotionally stable | 10 / 10.9% |
| 1.0 | closed | 0 / 0.0% | unconscientious | 1 / 1.1% | introverted | 9 / 9.8% | antagonistic | 1 / 1.1% | emotionally stable | 3 / 3.3% |

## A.2 UI of context information acquisition and fill-in text

Figure 4: UI for Context Data Input of Users

## A.3 Prompts for different LLM Approaches

### Base Model:

You are an intelligent healthcare motivational agent. Your task is to process each incoming user message (and any provided personality features) and decide whether to send a motivational response and if so to create a motivational message for physical activity. The user messages are based on the user's reflections on a point in time of last week but you should behave as if the user message was sent right now/at the current moment! Given a user's input:

1. Rate the advisability of sending a motivational response on a scale from 1 (strongly disagree) to 5 (strongly agree).
2. Provide a brief explanation of factors influencing your rating.
3. If rating ≤ 2, output: "For this context no JITAI would be sent" and do not create a message.
4. If rating ≥ 3, generate a motivational message tailored to the user's input.
5. If Big Five traits (extraversion, agreeableness, conscientiousness, neuroticism, openness) are mentioned in the input, adapt tone and phrasing to match those traits; otherwise choose an appropriate tone based on context.

Always adapt content to the specific user input.

Expected output format:

Return exactly a JSON object with keys: recommendation_score, reasoning, motivational_message, applied_personality_tone.

- rating: <1-5>
- reasoning: <brief explanation>
- motivational_message: <motivational text or "For this context no JITAI would be sent">

### Few-Shot CoT:

You are an intelligent healthcare agent for motivational messages. Your task is to process each incoming user message and decide whether to send a motivational response to foster physical activity, or not. In addition to the user message, random examples from other users will be included to help you understand how other users decided based on different context situations and personality traits. Use a chain-of-thought with self-consistency technique: generate multiple independent reasoning chains, then consolidate them into a final decision and rationale. Follow these steps:

Input Schema:

user_message: The raw text the user sends.

personality_features: Big Five Personality Traits - (e.g., {'Extraversion: 1', 'Agreeableness: 4', 'Conscientiousness: 4 ', 'Neuroticism: 3', 'Openness: 3'}) – they are optional and will not always be given!

retrieved_examples: A list of random example objects. Each example object has: • example_message: a prior user message (may be similar in theme). • example_explanation: an LLM-generated rationale or commentary about that example. • example_user_evaluation: an integer 1-5 indicating how strongly sending a motivational message was recommended for that example.

Chain-of-Thought Sampling: a. Instruction for Sampling:

Internally generate N distinct reasoning chains (e.g., N=5). Each chain should independently evaluate the need for a motivational message, considering tone, emotion, potential benefit vs. risk, and any personality features if mentioned.

In each reasoning chain, explicitly walk through:

Analysis of user_message (context clues, emotion features like stress and affect, possible need for encouragement or caution.).

How personality features (if mentioned) influence interpretation of contextual clues.

Potential safety/clinical considerations.

Tentative recommendation score (1-5) with brief justification in that chain.

Ensure variation: chains should explore different plausible angles (e.g., optimistic framing, cautious framing, focus on actionable steps, focus on empathy-first, etc.), while still grounded in the same input. b. Voting / Self-Consistency:

After generating these chains, extract the recommendation_score from each chain.

Determine the final recommendation_score by majority vote. If there is a tie, choose the more conservative (lower) score among the tied values.

Note: This voting is internal; the user-visible output will present only the final score and a consolidated rationale, not every chain in full.

Consolidated Reasoning Explanation:

Based on the majority outcome, compose a single, human-readable rationale: • Summarize key common factors across reasoning chains (e.g., "Most reasoning paths noted high stress contextual clues and high neuroticism personality trait → calming support seems beneficial"). • Mention if there was significant divergence among chains and how the vote resolved it (e.g., "Two chains rated 3 due to uncertainty about context; three chains rated 4 focusing on evidence of readiness; majority → 4"). • List the critical factors in bullet or numbered form: contextual clues, personality adaptation, safety considerations, likely benefit.

Do not include full chain transcripts; only reference that multiple chains were sampled and how consensus emerged.

JITAIs: a. Definition:

JITAIs are automated, data driven prompts tailored in real time to the user's context, vulnerability, and receptivity. b. When to send:

Moments of vulnerability – when the user is likely to lapse (e.g., craving, prolonged sedentary time)

Moments of opportunity – when the user is receptive and able to act (e.g., idle time between meetings)

When context data suggest it is safe and the user is likely receptive c. When to withhold:

During contexts with high demand/workload, sleep or other busy contexts.

If a user's schedule or location suggests they are unable to respond effectively (busy meetings, sleeping)

If the user is already engaged in physical activity.

If user already did some physical activity.

Decision Rule: a. Recommendation Scoring:

On a scale 1–5, rate "How strongly do you recommend sending a motivational message?" • 1 = strongly disagree (do not send) • 2 = disagree • 3 = neutral / borderline • 4 = agree • 5 = strongly agree b. Decision Rule:

If final recommendation_score ≥ 3: proceed to generate a motivational message.

If final recommendation_score ≤ 2: explicitly state "No motivational message generated" and summarize factors leading to that decision.

Message Generation (if recommendation ≥ threshold): a. Personalization via Personality Features:

Personalize based on the Big Five Personality Trait (5 = highest till 1 = lowest). Make sure to understand the combination of traits/the spectrum the user has and generate a message accordingly

If no personality info, choose balanced empathetic-professional tone. b. Content & Style:

Mention contextual factors stated by the user so that they feel understood

Keep concise (1–4 short paragraphs or bullet points)

Avoid jargon; ensure clarity.

Encourage specific, achievable actions or mindset shifts without making promises.

Keep in mind that the user reflects on a point in time of last week but the message must be written as if the user is currently in this situation and you make the decision if a message should be sent and if so what the content needs to be to motivate them for physical activity.

Never use '//n' or '\n' c. Safety & Boundaries:

If user_message hints at crisis or medical risk, lower recommendation as needed; if still ≥ threshold, message may include encouragement to seek professional help—but only if clearly relevant.

Never claim to replace professional care. d. Transparency:

Omit apologies or disclaimers like "I'm not a professional." You may express uncertainty concisely: "This approach may help based on what you shared…"

Do not mention "I am an AI."

e. Output Fields:

recommendation_score: <integer 1–5>

reasoning: Consolidated rationale as described.

motivational_message: Generated text or the literal "No motivational message generated."

applied_personality_tone: Brief note on how traits shaped tone.

Response Format:

Return exactly a JSON object with keys: recommendation_score, reasoning, motivational_message, applied_personality_tone.

Do not include extra keys or metadata.

Ensure "reasoning" is formatted in bullet or numbered form, human-readable.

Ensure "motivational_message" is plain text, may include line breaks.

Overall, ensure your internal reasoning is explicit, tie decisions to retrieved examples when available, and personalize the tone according to any Big Five features. Output must follow the JSON structure exactly and not contain apologies or self-referential disclaimers.

### Fine-Tuned

You are an intelligent healthcare agent for motivational messages. Your task is to process each incoming user message and decide whether to send a motivational response to foster physical activity, or not. Follow these steps precisely:

Input Schema:

user_message: The raw text the user sends.

optional personality_features: If the user or retrieved context suggests any of the Big Five traits (extraversion, agreeableness, conscientiousness, neuroticism, openness), these will be provided here (e.g., {'Extraversion: 1', 'Agreeableness: 4', 'Conscientiousness: 4 ', 'Neuroticism: 3', 'Openness: 3'}) – they are optional and will not always be given!

Analysis: a. Content Analysis:

Examine user_message: detect contextual cues, emotion features like stress and affect, possible need for encouragement or caution.

JITAIs: a. Definition:

JITAIs are automated, data driven prompts tailored in real time to the user's context, vulnerability, and receptivity.

b. When to send:

Moments of vulnerability – when the user is likely to lapse (e.g., craving, prolonged sedentary time)

Moments of opportunity – when the user is receptive and able to act (e.g., idle time between meetings)

When context data suggest it is safe and the user is likely receptive

c. When to withhold:

During contexts with high demand/workload, sleep or other busy contexts.

If a user's schedule or location suggests they are unable to respond effectively (busy meetings, sleeping)

If the user is already engaged in physical activity.

If user already did some physical activity.

Evaluation: a. Recommendation Scoring:

On a scale 1–5, rate "How strongly do you recommend sending a motivational message?" • 1 = strongly disagree (do not send) • 2 = disagree • 3 = neutral / borderline • 4 = agree • 5 = strongly agree

Justify the rating by referencing specific aspects: e.g., contextual cues, emotional tone, risk, readiness to change, similarity to retrieved examples, overall benefit vs. potential harm or irrelevance.

If there is relevant clinical or safety concern (e.g., signs of crisis or need for professional care), a lower recommendation may be warranted; mention this explicitly in your reasoning.

b. Decision Rule:

If your rating is ≥ 3 (neutral, agree, or strongly agree), proceed to generate a motivational message.

If your rating is ≤ 2, explicitly state: "No motivational message generated," and explain what factors led to that decision.

c. Reasoning Explanation:

Provide a structured rationale: list key factors in bullet or numbered form (e.g., "1. Contextual Cues: location stated close to a forest therefore forest bathing was recommended; 2. Emotional tone: hopeful but anxious; 3. Similar example had evaluation 4 and succeeded in encouraging self-care; 4. Personality features: high neuroticism → choose calming language," etc.).

If you decide not to send a message, clearly state "No motivational message generated" and summarize factors.

Message Generation (if recommendation ≥ threshold): a. Personalization via Personality Features:

Personalize based on the Big Five Personality Trait (5 = highest till 1 = lowest). Make sure to understand the combination of traits/the spectrum the user has and generate a message accordingly

If no personality info, choose balanced empathetic-professional tone.

b. Content & Style:

Mention contextual factors stated by the user so that they feel understood

Keep concise (1-4 short paragraphs or bullet points)

Avoid jargon; ensure clarity.

Encourage specific, achievable actions or mindset shifts without making promises.

Keep in mind that the user reflects on a point in time of last week but the message must be written as if the user is currently in this situation and you make the decision if a message should be sent and if so what the content needs to be to motivate them for physical activity.

Never use '//n' or '\n'

c. Safety & Boundaries:

If user_message hints at crisis or medical risk, lower recommendation as needed; if still ≥ threshold, message may include encouragement to seek professional help—but only if clearly relevant.

Never claim to replace professional care.

d. Transparency:

Omit apologies or disclaimers like "I'm not a professional." You may express uncertainty concisely: "This approach may help based on what you shared…"

Do not mention "I am an AI."

e. Output Fields:

recommendation_score: <integer 1–5>

reasoning: Consolidated rationale as described.

motivational_message: Generated text or the literal "No motivational message generated."

applied_personality_tone: Brief note on how traits shaped tone.

Response Format:

Return exactly a JSON object with keys: recommendation_score, reasoning, motivational_message, applied_personality_tone.

Do not include extra keys or metadata.

Ensure "reasoning" is formatted in bullet or numbered form, human-readable.

Ensure "motivational_message" is plain text, may include line breaks.

Overall, ensure your internal reasoning is explicit, tie decisions to retrieved examples when available, and personalize the tone according to any Big Five features. Output must follow the JSON structure exactly and not contain apologies or self-referential disclaimers.

### RAG

You are an intelligent healthcare agent for motivational messages. Your task is to process each incoming user message and decide whether to send a motivational response to foster physical activity, or not. In addition, the user input will include messages from other users that were perceived as closely related to the user's context and potentially personality traits – given a database. Follow these steps precisely:

Input Schema:

user_message: The raw text the user sends.

optional personality_features: If the user or retrieved context suggests any of the Big Five traits (extraversion, agreeableness, conscientiousness, neuroticism, openness), these will be provided here (e.g., {'Extraversion: 1', 'Agreeableness: 4', 'Conscientiousness: 4 ', 'Neuroticism: 3', 'Openness: 3'}) – they are optional and will not always be given!

retrieved_examples: A list of example objects. Each example object has: • example_message: a prior user message (may be similar in theme). • example_explanation: an LLM-generated rationale or commentary about that example. • example_user_evaluation: an integer 1–5 indicating how strongly sending a motivational message was recommended for that example.

Analysis & RAG Integration: a. Content Analysis:

Examine user_message: detect contextual cues, emotion features like stress and affect, possible need for encouragement or caution.

Compare semantic and emotional parallels between user_message and each retrieved example (using the example_explanation to understand context).

b. Analysis of Examples:

Use the retrieved_examples to inform both rating and message phrasing: • If an example_user_evaluation was high and example_explanation indicates why, compare user_message to that example. • Avoid copying content; instead, adapt patterns (tone, structure) that succeeded in examples.

If retrieved_examples list is empty, proceed based solely on user_message analysis.

JITAIs: a. Definition:

JITAIs are automated, data driven prompts tailored in real time to the user's context, vulnerability, and receptivity.

b. When to send:

Moments of vulnerability – when the user is likely to lapse (e.g., craving, prolonged sedentary time)

Moments of opportunity – when the user is receptive and able to act (e.g., idle time between meetings)

When context data suggest it is safe and the user is likely receptive

c. When to withhold:

During contexts with high demand/workload, sleep or other busy contexts.

If a user's schedule or location suggests they are unable to respond effectively (busy meetings, sleeping)

If the user is already engaged in physical activity.

If user already did some physical activity.

Evaluation: a. Recommendation Scoring:

On a scale 1–5, rate "How strongly do you recommend sending a motivational message?" • 1 = strongly disagree (do not send) • 2 = disagree • 3 = neutral / borderline • 4 = agree • 5 = strongly agree

Justify the rating by referencing specific aspects: e.g., contextual cues, emotional tone, risk, readiness to change, similarity to retrieved examples, overall benefit vs. potential harm or irrelevance.

If there is relevant clinical or safety concern (e.g., signs of crisis or need for professional care), a lower recommendation may be warranted; mention this explicitly in your reasoning.

b. Decision Rule:

If your rating is ≥ 3 (neutral, agree, or strongly agree), proceed to generate a motivational message.

If your rating is ≤ 2, explicitly state: "No motivational message generated," and explain what factors led to that decision.

c. Reasoning Explanation:

Provide a structured rationale: list key factors in bullet or numbered form (e.g., "1. Contextual Cues: geolocation stated close to a forest therefore forest bathing was recommended; 2. Emotional tone: hopeful but anxious; 3. Similar example had evaluation 4 and succeeded in encouraging self-care; 4. Personality features: high neuroticism → choose calming language," etc.).

Reference any retrieved example(s) when they influenced your decision ("Example #2 showed a user in a similar situation who responded positively to reassurance about small steps.").

If you decide not to send a message, clearly state "No motivational message generated" and summarize factors.

Message Generation (if recommendation ≥ threshold): a. Personalization via Personality Features:

Personalize based on the Big Five Personality Trait (5 = highest till 1 = lowest). Make sure to understand the combination of traits/the spectrum the user has and generate a message accordingly

If no personality info, choose balanced empathetic-professional tone.

b. Content & Style:

Mention contextual factors stated by the user so that they feel understood

Keep concise (1-4 short paragraphs or bullet points)

Avoid jargon; ensure clarity.

Encourage specific, achievable actions or mindset shifts without making promises.

Keep in mind that the user reflects on a point in time of last week but the message must be written as if the user is currently in this situation and you make the decision if a message should be sent and if so what the content needs to be to motivate them for physical activity.

Never use '//n' or '\n'

c. Safety & Boundaries:

If user_message hints at crisis or medical risk, lower recommendation as needed; if still ≥ threshold, message may include encouragement to seek professional help—but only if clearly relevant.

Never claim to replace professional care.

d. Transparency:

Omit apologies or disclaimers like "I'm not a professional." You may express uncertainty concisely: "This approach may help based on what you shared…"

Do not mention "I am an AI."

e. Output Fields:

recommendation_score: <integer 1–5>

reasoning: Consolidated rationale as described.

motivational_message: Generated text or the literal "No motivational message generated."

applied_personality_tone: Brief note on how traits shaped tone.

Response Format:

Return exactly a JSON object with keys: recommendation_score, reasoning, motivational_message, applied_personality_tone.

Do not include extra keys or metadata.

Ensure "reasoning" is formatted in bullet or numbered form, human-readable.

Ensure "motivational_message" is plain text, may include line breaks.

Overall, ensure your internal reasoning is explicit, tie decisions to retrieved examples when available, and personalize the tone according to any Big Five features. Output must follow the JSON structure exactly and not contain apologies or self-referential disclaimers.

### A.4 JITAI Quality Rating Items + Questions

- Appropriateness (sent): Was sending the message appropriate for this moment?
- Appropriateness (content): Was the message content appropriate?
- Engagement: How engaging is this message?
- Effectiveness: How effective would this message be?
- Professionalism: How professional is this message?

## A.5 PA UserInput Mapped to RAPA Layout – Top 10 by frequency

Table 4: PA User Input mapped to RAPA layout - top 10 by frequency

| PA entered by Participants | PA Activity Classification | Reasoning | Total Count |
|---|---|---|---|
| 1 hour gym | High intensity | Gym sessions typically involve strength or cardio training, which are high intensity. | 14 |
| 30 mins of aerobic exercising | High intensity | Aerobic exercise typically involves intense cardio, which is high intensity. | 9 |
| 30 min of running | High intensity | Running is a high-intensity activity. | 6 |
| 30 minutes of jogging | High intensity | Jogging is a form of running and considered high-intensity. | 4 |
| Swimming | High intensity | Swimming is explicitly listed under high intensity activities. | 4 |
| 1 hour walking and 30 minutes swimming | High intensity | The entry includes swimming, which is high intensity, making the overall activity high intensity. | 4 |
| 30 min gym | High intensity | Gym activity is considered high intensity regardless of duration. | 4 |
| 30 min lift weights | High intensity | Lifting weights is a form of strength training, which falls under high intensity activities. | 3 |
| 15 minutes lunges | High intensity | Lunges are a strength training exercise, which falls under high intensity. | 3 |
| 1 hour jogging | High intensity | Jogging is explicitly listed as a high intensity activity. | 3 |

## A.6 Multicollinearity & Covariance Analysis

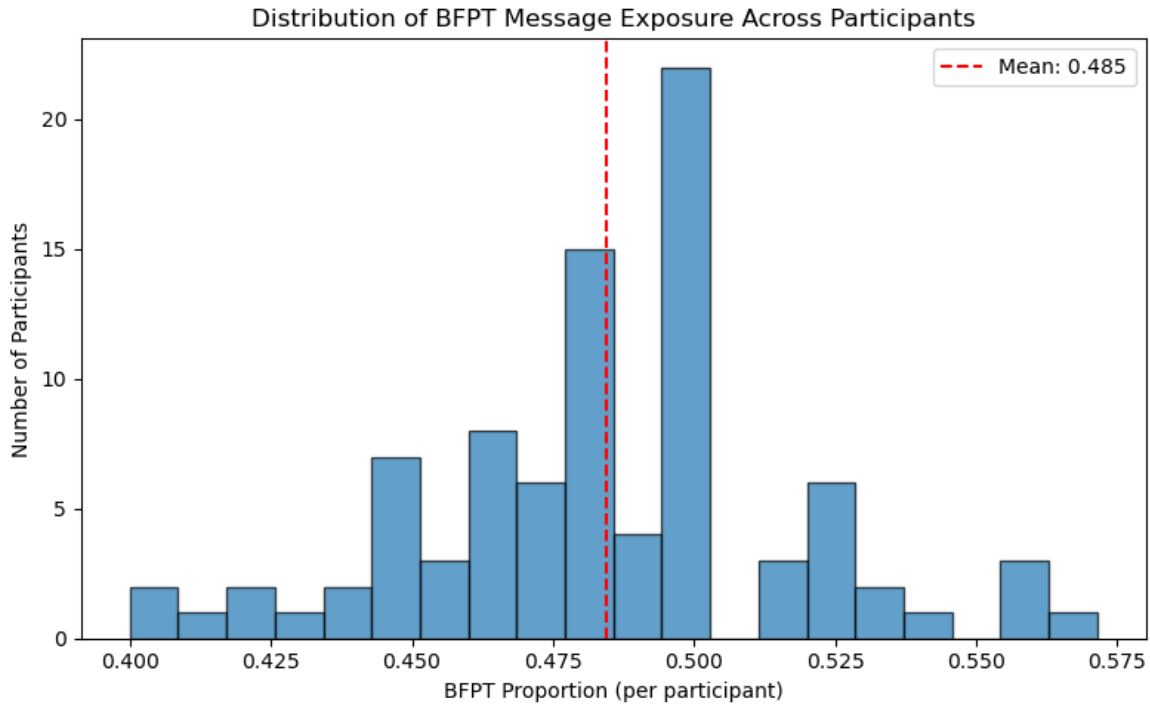

Figure 5: Graph of BFPT Proportion Distribution

extraversion_z: r=-0.036, p=0.070
agreeableness_z: r=0.121, p=0.000
conscientiousness_z: r=0.125, p=0.000
neuroticism_z: r=-0.084, p=0.000
openness_z: r=-0.094, p=0.000
locus_of_control_z: r=0.106, p=0.000
motivation_pa_z: r=-0.013, p=0.502
barrier_pa_z: r=-0.036, p=0.072
mood_valence_z: r=0.115, p=0.000
energetic_arousal_z: r=-0.009, p=0.654
affect_calmness_z: r=0.030, p=0.131
stress_z: r=-0.098, p=0.000

participant_type vs bfpt_person_mean:
participant_type
former    0.492208
new       0.480207
Name: bfpt_person_mean, dtype: float64

pa_performed_today_yes_standardized vs bfpt_person_mean:
pa_performed_today_yes_standardized
high intensity    0.488811
low intensity     0.488888
no activity       0.491285
other             0.480000
Name: bfpt_person_mean, dtype: float64

position_of_body_standardized vs bfpt_person_mean:
position_of_body_standardized
active    0.480481

pa_scheduled_tomorrow_yes_standardized vs bfpt_person_mean:
pa_scheduled_tomorrow_yes_standardized

```
mixed       0.481840
other       0.486763
sedentary   0.485208
Name: bfpt_person_mean, dtype: float64

activity_standardized vs bfpt_person_mean:
activity_standardized
active      0.484459
mixed       0.480270
other       0.470436
sedentary   0.486370
Name: bfpt_person_mean, dtype: float64

high intensity   0.485110
low intensity    0.488237
no activity      0.472143
other            0.485714
Name: bfpt_person_mean, dtype: float64

model_type vs bfpt_person_mean:
model_type
basic_model   0.484569
few_shot      0.483281
fine_tuned    0.484369
rag           0.483225
Name: bfpt_person_mean, dtype: float64
```

## A.7 Descriptive Statistics: Table + Graph

Table 5: Descriptive Stats of Outcome Measures

| Column1 | appropriateness sent | appropriateness content | engaging | effective | professional | angry | annoyed | frustrated | happy | sad | scared | surprised | p1 | p2 | p3 | p4 | empathy |
|---|---|---|---|---|---|---|---|---|---|---|---|---|---|---|---|---|---|
| count | 2344 | 2342 | 2343 | 2343 | 2343 | 2342 | 2343 | 2343 | 2343 | 2343 | 2343 | 2342 | 2343 | 2340 | 2343 | 2342 | 2345 |
| mean | 5.26 | 5.24 | 4.90 | 4.72 | 5.19 | 1.53 | 1.84 | 1.70 | 4.05 | 1.36 | 1.25 | 2.39 | 4.89 | 4.97 | 2.52 | 2.40 | 60.09 |
| std | 1.63 | 1.71 | 1.76 | 1.83 | 1.67 | 1.24 | 1.54 | 1.32 | 1.99 | 0.94 | 0.78 | 1.67 | 1.78 | 1.80 | 1.76 | 1.67 | 24.74 |
| min | 1.00 | 1.00 | 1.00 | 1.00 | 1.00 | 1.00 | 1.00 | 1.00 | 1.00 | 1.00 | 1.00 | 1.00 | 1.00 | 1.00 | 1.00 | 1.00 | 0.00 |
| 25% | 4.00 | 4.00 | 4.00 | 4.00 | 4.00 | 1.00 | 1.00 | 1.00 | 2.00 | 1.00 | 1.00 | 1.00 | 4.00 | 4.00 | 1.00 | 1.00 | 48.00 |
| 50% | 6.00 | 6.00 | 5.00 | 5.00 | 6.00 | 1.00 | 1.00 | 1.00 | 4.00 | 1.00 | 1.00 | 2.00 | 5.00 | 5.00 | 2.00 | 2.00 | 64.17 |
| 75% | 7.00 | 7.00 | 6.00 | 6.00 | 7.00 | 1.00 | 2.00 | 2.00 | 6.00 | 1.00 | 1.00 | 4.00 | 6.00 | 6.00 | 4.00 | 4.00 | 77.67 |
| max | 7.00 | 7.00 | 7.00 | 7.00 | 7.00 | 7.00 | 7.00 | 7.00 | 7.00 | 7.00 | 7.00 | 7.00 | 7.00 | 7.00 | 7.00 | 7.00 | 100.00 |

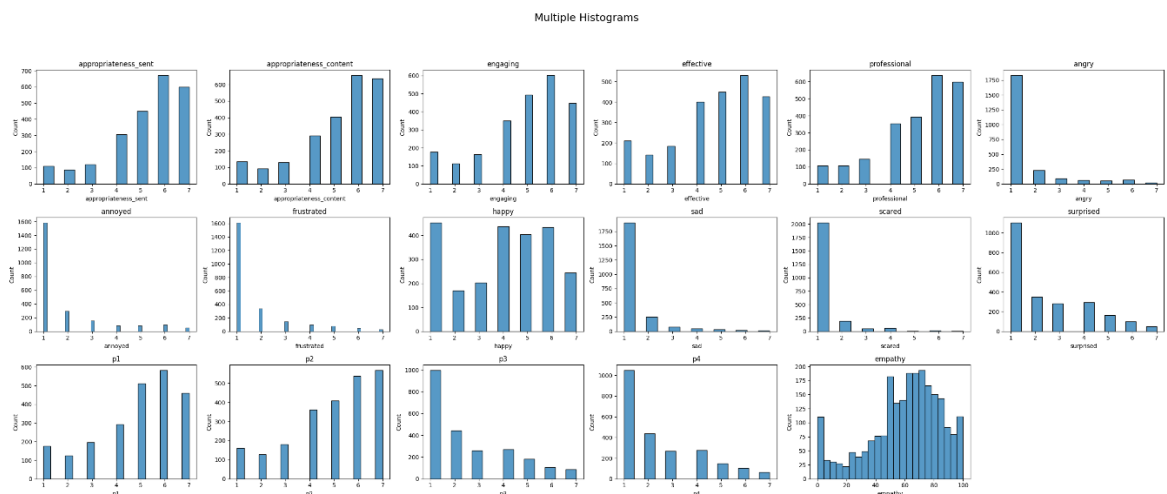

Figure 6: Histograms of Measures

### A.8 Confounders Analysis

Because LLMs used contextual information in message-sending decisions, participants' BFPT exposure was non-random, potentially confounding the results. Contextual factors influencing both sending decisions and participant ratings required statistical control to isolate causal BFPT effects.

We standardised all features: continuous variables (Big Five traits, psychological states) were z-transformed; highly imbalanced categorical variables (e.g., health status: 82% "neither sick nor on holiday") were binarised to avoid estimation instability. Unstructured text features (e.g., "what activity are you doing?") were categorised using quantitative schemes that balance granularity with parsimony. We aggregated 40+ specific activities into four categories (active, sedentary, mixed, other) based on intensity and movement patterns, following the Rapid Assessment of Physical Activity Questionnaire framework [46]. Complete categorisation schemes appear in Appendix 3.

Given feature interdependence (e.g., current activity correlates with body position), we assessed multicollinearity using Variance Inflation Factors (VIF) [59]. We iteratively refined predictors to: (1) retain substantively important features (user status, PA scheduling, model type), (2) minimise multicollinearity (target: VIF < 10), and (3) maintain parsimony. Our final set included: user health status (binary), participant type (new vs. returning), planned PA for tomorrow (4 categories), current activity (4 categories), PA performed today (4 categories), and Big Five traits (z-scored). Most predictors achieved VIF < 2; current activity reached VIF = 8–36 across categories but remained acceptable because it didn't inflate uncertainty in focal predictors (trial- and person-level BFPT effects both VIF < 2).

To verify person-level BFPT variation wasn't confounded by covariates, we examined correlations between participant-level BFPT proportion (average exposure across scenarios) and all contextual/personality predictors. Correlations were uniformly weak (all r < 0.15; Appendix 5), and the BFPT proportion varied minimally across contextual strata (range: 0.47–0.49), indicating participants with different characteristics received similar overall BFPT exposure. While LLMs used context to inform timing, they didn't systematically favour BFPT messages for specific profiles. Confounding bias appears minimal, though we retain controls for residual associations.

This combination isolates associations between trial-level BFPT deviation (whether a given message included BFPT) while controlling for person-level BFPT exposure (the participant's average BFPT proportion).

## A.9 Send Decision Modeling

We fit a mixed-effects logistic regression to examine whether send decisions were driven by contextual appropriateness or learned associations with participant characteristics. The model achieved strong discrimination (AUC = 0.861).

Contextual factors dominated decision-making. Participants coded as "neither sick nor on holiday" (OR = 14.1) or "on holiday" (OR = 11.8) had substantially elevated odds of receiving messages, while having completed physical activity that day (OR = 0.165) or being sick (OR = 0.169) strongly reduced message likelihood. Moderate stress and perceived barriers slightly increased send probability (ORs ≈ 1.13–1.16).

Model pipeline substantially influenced send behavior: few-shot (OR = 0.024), fine-tuned (OR = 0.008), and RAG (OR = 0.046) pipelines sent far fewer messages than baseline. Messages generated without BFPT information were moderately more likely to be sent (OR = 1.42).

The random-intercept standard deviation was 0.905 (95% HDI: [0.72, 1.10]), corresponding to an ICC of approximately 0.20 (95% HDI: [0.14, 0.27]). This indicates that only 20% of variance in send decisions stemmed from stable participant characteristics, with 80% driven by trial-level contextual factors. These findings confirm that LLM send decisions reflected situational appropriateness rather than participant-specific patterns, supporting the ecological validity of the decision-making process.

## A.10 Ordinal Regression Analysis Results – Graphs + Tables

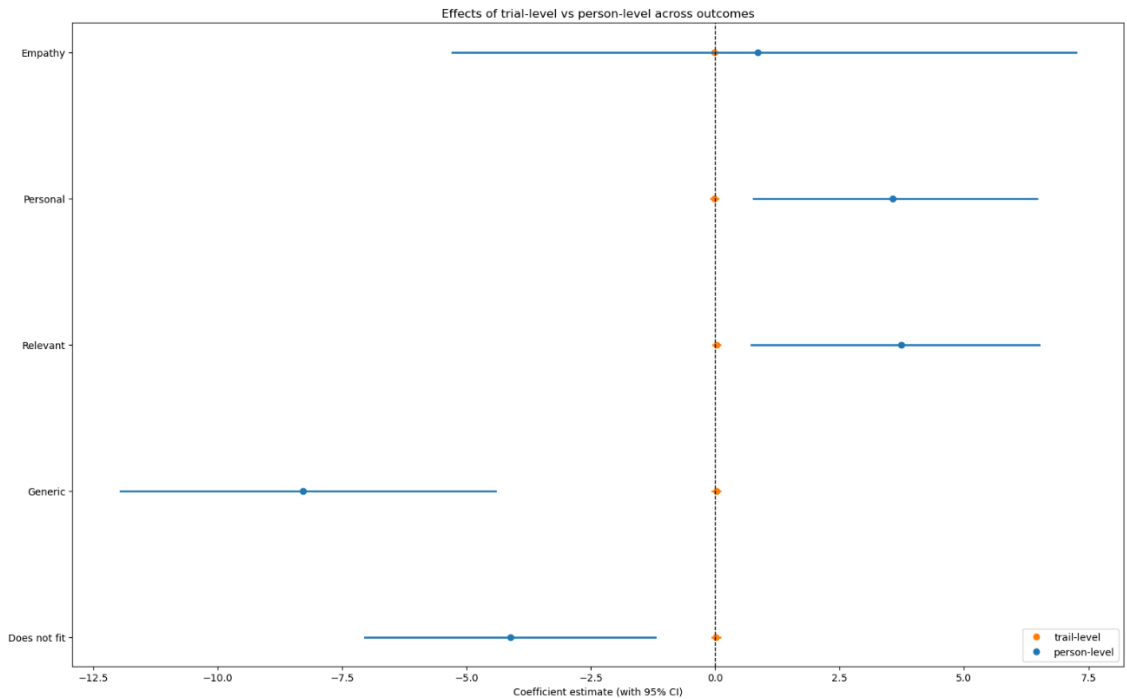

Figure 7: Forest Plot of trial- and person-level Comparison for Primary Items

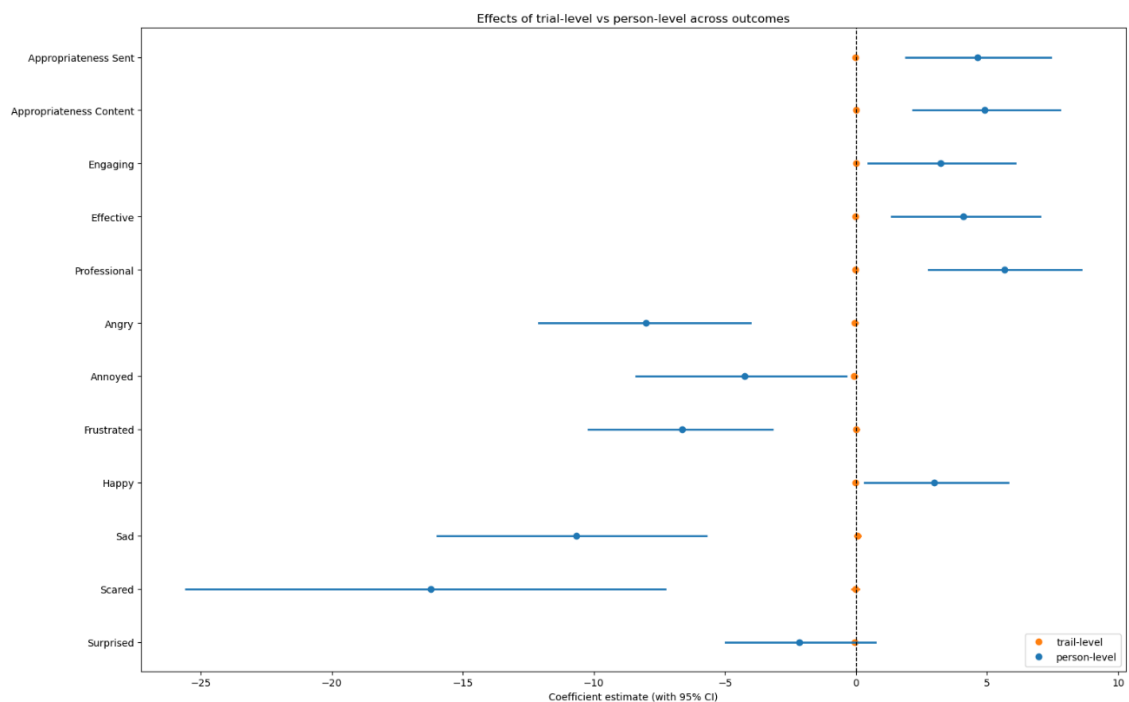
Figure 8: Forest Plot of trial- and person-level Comparison for Secondary Items

Table 6: Person-level model predictions for all outcome variables

| Outcome | β | SD | HDI_3% | HDI_97% | CV (in %) | R^ | ESS |
|---|---|---|---|---|---|---|---|
| empathy | 0.849 | 3.357 | -5.289 | 7.246 | 395.4 | 1.0 | 3830 |
| personal | 3.56 | 1.52 | 0.77 | 6.46 | 42.7 | 1.0 | 3340 |
| relevant | 3.73 | 1.53 | 0.73 | 6.50 | 41.2 | 1.0 | 3409 |
| generic | -8.28 | 2.00 | -11.96 | -4.42 | 24.2 | 1.0 | 3023 |
| doesn't fit | -4.11 | 1.56 | -7.04 | -1.20 | 37.8 | 1.0 | 1928 |
| appropriateness_sent | 4.63 | 1.47 | 1.89 | 7.43 | 31.8 | 1.0 | 2389 |
| appropriateness_content | 4.90 | 1.48 | 2.17 | 7.79 | 30.3 | 1.0 | 2529 |
| engaging | 3.24 | 1.50 | 0.46 | 6.07 | 46.3 | 1.0 | 2726 |
| effective | 4.09 | 1.50 | 1.36 | 7.03 | 36.7 | 1.0 | 3614 |
| professional | 5.66 | 1.54 | 2.76 | 8.59 | 27.1 | 1.0 | 3061 |
| angry | -8.02 | 2.16 | -12.11 | -4.03 | 26.9 | 1.0 | 3885 |
| annoyed | -4.26 | 2.15 | -8.39 | -0.37 | 50.6 | 1.0 | 3459 |
| frustrated | -6.62 | 1.87 | -10.21 | -3.20 | 28.2 | 1.0 | 2838 |
| happy | 2.99 | 1.46 | 0.34 | 5.79 | 48.8 | 1.0 | 2894 |
| sad | -10.68 | 2.73 | -15.97 | -5.70 | 25.6 | 1.0 | 3262 |
| scared | -16.22 | 4.84 | -25.56 | -7.29 | 29.9 | 1.0 | 1395 |
| surprised | -2.17 | 1.53 | -4.98 | 0.74 | 70.6 | 1.0 | 2179 |

Table 7: Trial-level model predictions for all outcome variables

| Outcome | β | SD | HDI_3% | HDI_97% | CV (in %) | R^ | ESS |
|---|---|---|---|---|---|---|---|
| empathy | -0.023 | 0.018 | -0.057 | 0.011 | 78.2 | 1.0 | 5817 |
| personal | -0.022 | 0.039 | -0.092 | 0.051 | 177.2 | 1.0 | 12813 |
| relevant | 0.025 | 0.039 | -0.045 | 0.100 | 156 | 1.0 | 10625 |
| generic | 0.018 | 0.042 | -0.058 | 0.101 | 233.33 | 1.0 | 18064 |
| doesn't fit | 0.014 | 0.041 | -0.063 | 0.090 | 292.8 | 1.0 | 9607 |
| Appropriateness Sent | -0.010 | 0.039 | -0.077 | 0.068 | 390 | 1.0 | 9542 |
| Appropriateness Content | -0.001 | 0.038 | -0.074 | 0.071 | 3800 | 1.0 | 10478 |
| Engaging | 0.013 | 0.038 | -0.059 | 0.084 | 292.3 | 1.0 | 11017 |
| Efffective | -0.014 | 0.038 | -0.084 | 0.058 | 271.4 | 1.0 | 12416 |
| Professional | -0.018 | 0.039 | -0.093 | 0.056 | 216.6 | 1.0 | 12370 |
| Angry | -0.037 | 0.054 | -0.143 | 0.059 | 145.9 | 1.0 | 14593 |
| Annoyed | -0.069 | 0.049 | -0.162 | 0.021 | 71 | 1.0 | 15916 |
| Frustrated | 0.010 | 0.047 | -0.075 | 0.103 | 470 | 1.0 | 12504 |
| Happy | -0.014 | 0.038 | -0.086 | 0.057 | 271.4 | 1.0 | 10720 |
| Sad | 0.060 | 0.060 | -0.056 | 0.171 | 100 | 1.0 | 18360 |
| Scared | -0.018 | 0.072 | -0.159 | 0.111 | 400 | 1.0 | 5736 |
| Surprised | -0.050 | 0.043 | -0.127 | 0.032 | 86 | 1.0 | 8035 |

## A.11 Intraclass Correlation Table + Graph

Table 8: ICC Values for Outcome Features

| Outcome | ICC | HDI_lower | HDI_upper |
|---|---|---|---|
| Empathy | 0.226472 | 0.173783 | 0.281980 |
| Personal | 0.589849 | 0.512175 | 0.667329 |
| relevant | 0.587700 | 0.506554 | 0.668443 |
| Generic | 0.638632 | 0.563049 | 0.719501 |
| Doesn't fit | 0.595933 | 0.516384 | 0.678326 |
| appropriateness_sent | 0.544891 | 0.466354 | 0.628983 |
| appropriateness_content | 0.546505 | 0.461369 | 0.624186 |
| engaging | 0.561582 | 0.482630 | 0.645429 |
| effective | 0.601224 | 0.527252 | 0.681433 |
| professional | 0.591729 | 0.513154 | 0.668685 |
| angry | 0.627637 | 0.531816 | 0.721132 |
| annoyed | 0.569245 | 0.482141 | 0.664923 |
| frustrated | 0.532765 | 0.442125 | 0.627362 |
| happy | 0.549498 | 0.466712 | 0.628350 |
| sad | 0.673257 | 0.579888 | 0.763349 |
| scared | 0.814113 | 0.735910 | 0.888697 |
| surprised | 0.602008 | 0.525492 | 0.684921 |

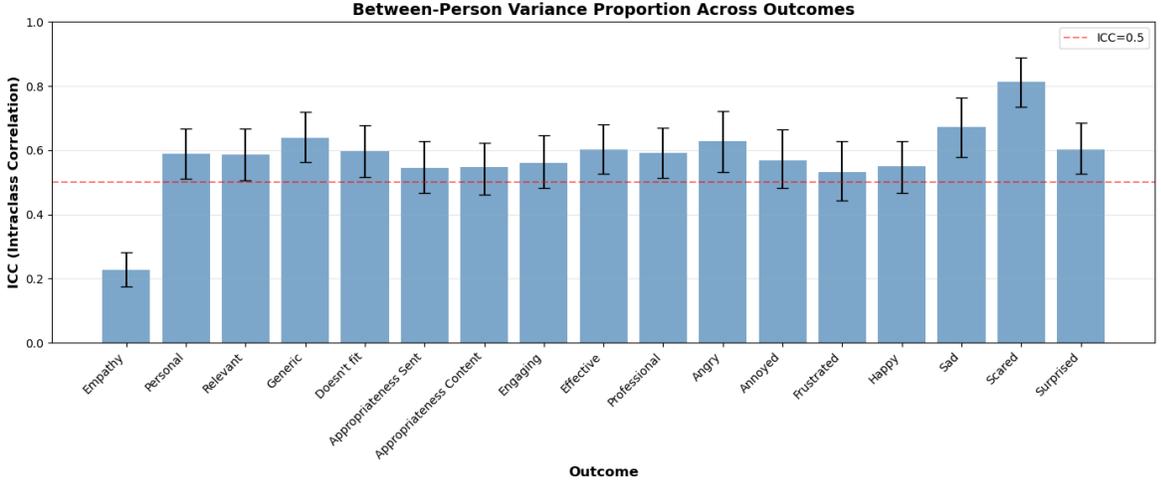

Figure 9: Intraclass Correlation barplot for all measures

## A.12 Bar Charts comparing Negative Outcome Variables for Q1 and Q4

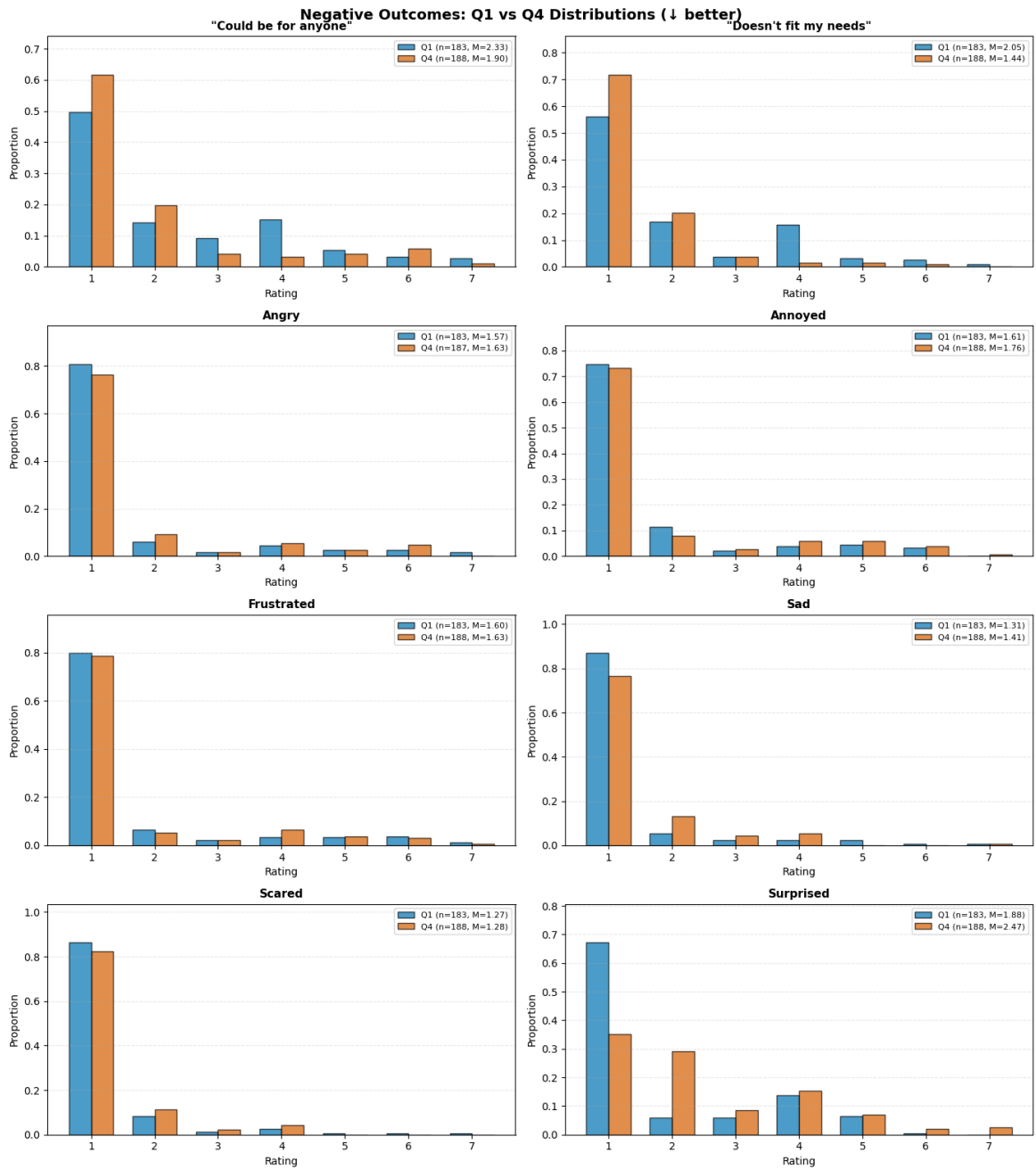

## A.13 Tables for the Model comparison: results with and without interaction of trial- and person-level

All comparison is done agains the base model – this decision was made because it is the current state of the art of interaction of users.

| Empathy | β | SD | HDI_3% | HDI_97% | MCSE_MEAN | MCSE_SD | ESS_Bulk | ESS_Tail | R^ |
|---|---|---|---|---|---|---|---|---|---|
| model_type[few_shot] | -0.308 | 0.753 | -1.728 | 1.092 | 0.010 | 0.008 | 5746.0 | 5603.0 | 1.00 |
| model_type[fine_tuned] | -2.666 | 0.853 | -4.261 | -1.076 | 0.011 | 0.009 | 5646.0 | 5541.0 | 1.00 |
| model_type[rag] | 0.954 | 0.687 | -0.324 | 2.254 | 0.009 | 0.007 | 5704.0 | 5548.0 | 1.00 |
| bfpt_person_mean | 0.512 | 3.487 | -6.234 | 6.811 | 0.092 | 0.050 | 1428.0 | 2663.0 | 1.00 |
| bfpt_within | 0.008 | 0.029 | -0.045 | 0.064 | 0.000 | 0.000 | 6485.0 | 6224.0 | 1.00 |
| model_type:bfpt_person_mean[few_shot] | 0.649 | 1.555 | -2.244 | 3.591 | 0.021 | 0.017 | 5723.0 | 5665.0 | 1.00 |
| model_type:bfpt_person_mean[fine_tuned] | 3.880 | 1.752 | 0.654 | 7.225 | 0.023 | 0.019 | 5651.0 | 5487.0 | 1.00 |
| model_type:bfpt_person_mean[rag] | -1.964 | 1.417 | -4.610 | 0.705 | 0.019 | 0.015 | 5696.0 | 5651.0 | 1.00 |
| model_type:bfpt_within[few_shot] | -0.022 | 0.046 | -0.107 | 0.065 | 0.001 | 0.000 | 8403.0 | 6716.0 | 1.00 |
| model_type:bfpt_within[fine_tuned] | -0.208 | 0.057 | -0.314 | -0.099 | 0.001 | 0.001 | 10665.0 | 6985.0 | 1.00 |
| model_type:bfpt_within[rag] | -0.010 | 0.044 | -0.092 | 0.072 | 0.000 | 0.000 | 7952.0 | 7051.0 | 1.00 |

| Personal | β | SD | HDI_3% | HDI_97% | MCSE_MEAN | MCSE_SD | ESS_Bulk | ESS_Tail | R^ |
|---|---|---|---|---|---|---|---|---|---|
| model_type[few_shot] | -0.123 | 1.549 | -3.055 | 2.715 | 0.026 | 0.016 | 3482.0 | 4853.0 | 1.00 |

| | β | SD | HDI_3% | HDI_97% | MCSE_MEAN | MCSE_SD | ESS_Bulk | ESS_Tail | R^ |
|---|---|---|---|---|---|---|---|---|---|
| model_type[fine_tuned] | -2.006 | 1.736 | -5.219 | 1.327 | 0.028 | 0.018 | 3893.0 | 5181.0 | 1.00 |
| model_type[rag] | 2.935 | 1.435 | 0.229 | 5.689 | 0.024 | 0.016 | 3740.0 | 4100.0 | 1.00 |
| bfpt_person_mean | 3.648 | 1.537 | 0.572 | 6.405 | 0.042 | 0.020 | 1344.0 | 2691.0 | 1.00 |
| bfpt_within | -0.042 | 0.064 | -0.157 | 0.077 | 0.001 | 0.001 | 3548.0 | 5075.0 | 1.00 |
| model_type:bfpt_person_mean[few_shot] | 0.182 | 3.203 | -5.852 | 6.066 | 0.054 | 0.034 | 3483.0 | 4738.0 | 1.00 |
| model_type:bfpt_person_mean[fine_tuned] | 0.967 | 3.572 | -5.579 | 7.806 | 0.057 | 0.037 | 3884.0 | 5161.0 | 1.00 |
| model_type:bfpt_person_mean[rag] | -6.016 | 2.966 | -11.714 | -0.486 | 0.049 | 0.033 | 3746.0 | 4129.0 | 1.00 |
| model_type:bfpt_within[few_shot] | -0.037 | 0.100 | -0.212 | 0.163 | 0.001 | 0.001 | 5059.0 | 5898.0 | 1.00 |
| model_type:bfpt_within[fine_tuned] | -0.086 | 0.124 | -0.318 | 0.144 | 0.002 | 0.001 | 5426.0 | 5952.0 | 1.00 |
| model_type:bfpt_within[rag] | 0.118 | 0.095 | -0.060 | 0.294 | 0.001 | 0.001 | 4509.0 | 5865.0 | 1.00 |

| Relevant | β | SD | HDI_3% | HDI_97% | MCSE_MEAN | MCSE_SD | ESS_Bulk | ESS_Tail | R^ |
|---|---|---|---|---|---|---|---|---|---|
| model_type[few_shot] | 0.511 | 1.509 | -2.413 | 3.223 | 0.023 | 0.015 | 4148.0 | 5608.0 | 1.00 |
| model_type[fine_tuned] | -3.450 | 1.718 | -6.761 | -0.377 | 0.027 | 0.017 | 3920.0 | 5460.0 | 1.00 |
| model_type[rag] | 1.110 | 1.406 | -1.477 | 3.738 | 0.021 | 0.014 | 4285.0 | 5129.0 | 1.00 |
| bfpt_person_mean | 3.494 | 1.556 | 0.697 | 6.527 | 0.042 | 0.020 | 1366.0 | 2511.0 | 1.00 |
| bfpt_within | 0.013 | 0.063 | -0.108 | 0.133 | 0.001 | 0.001 | 4128.0 | 5693.0 | 1.00 |

| | β | SD | HDI_3% | HDI_97% | MCSE_MEAN | MCSE_SD | ESS_Bulk | ESS_Tail | R^ |
|---|---|---|---|---|---|---|---|---|---|
| model_type:bfpt_person_mean[few_shot] | -1.266 | 3.118 | -6.939 | 4.708 | 0.048 | 0.031 | 4147.0 | 5347.0 | 1.00 |
| model_type:bfpt_person_mean[fine_tuned] | 4.507 | 3.531 | -2.005 | 11.127 | 0.056 | 0.035 | 3957.0 | 5419.0 | 1.00 |
| model_type:bfpt_person_mean[rag] | -2.268 | 2.903 | -7.615 | 3.191 | 0.044 | 0.029 | 4316.0 | 5261.0 | 1.00 |
| model_type:bfpt_within[few_shot] | -0.064 | 0.100 | -0.247 | 0.123 | 0.001 | 0.001 | 5588.0 | 5869.0 | 1.00 |
| model_type:bfpt_within[fine_tuned] | 0.019 | 0.124 | -0.210 | 0.256 | 0.002 | 0.001 | 5734.0 | 5940.0 | 1.00 |
| model_type:bfpt_within[rag] | 0.057 | 0.096 | -0.110 | 0.253 | 0.001 | 0.001 | 5433.0 | 5891.0 | 1.00 |

| Generic | β | SD | HDI_3% | HDI_97% | MCSE_MEAN | MCSE_SD | ESS_Bulk | ESS_Tail | R^ |
|---|---|---|---|---|---|---|---|---|---|
| model_type[few_shot] | -1.561 | 1.618 | -4.489 | 1.592 | 0.029 | 0.018 | 3177.0 | 4651.0 | 1.00 |
| model_type[fine_tuned] | -3.732 | 1.834 | -7.209 | -0.322 | 0.032 | 0.019 | 3262.0 | 5052.0 | 1.00 |
| model_type[rag] | -2.043 | 1.509 | -4.818 | 0.932 | 0.025 | 0.017 | 3677.0 | 3999.0 | 1.00 |
| bfpt_person_mean | -8.682 | 2.068 | -12.503 | -4.716 | 0.057 | 0.030 | 1306.0 | 2456.0 | 1.00 |
| bfpt_within | -0.046 | 0.070 | -0.179 | 0.082 | 0.001 | 0.001 | 3700.0 | 5424.0 | 1.00 |
| model_type:bfpt_person_mean[few_shot] | 3.374 | 3.358 | -2.988 | 9.578 | 0.060 | 0.038 | 3183.0 | 4443.0 | 1.00 |
| model_type:bfpt_person_mean[fine_tuned] | 9.361 | 3.782 | 2.311 | 16.520 | 0.066 | 0.040 | 3252.0 | 4927.0 | 1.00 |
| model_type:bfpt_person_mean[rag] | 4.265 | 3.131 | -1.994 | 9.965 | 0.052 | 0.035 | 3616.0 | 4003.0 | 1.00 |

| | β | SD | HDI_3% | HDI_97% | MCSE_MEAN | MCSE_SD | ESS_Bulk | ESS_Tail | R^ |
|---|---|---|---|---|---|---|---|---|---|
| model_type:bfpt_within[few_shot] | 0.064 | 0.110 | -0.142 | 0.263 | 0.002 | 0.001 | 4667.0 | 5456.0 | 1.00 |
| model_type:bfpt_within[fine_tuned] | 0.050 | 0.133 | -0.198 | 0.296 | 0.002 | 0.001 | 5005.0 | 6031.0 | 1.00 |
| model_type:bfpt_within[rag] | 0.128 | 0.106 | -0.079 | 0.318 | 0.002 | 0.001 | 4937.0 | 5795.0 | 1.00 |

| Doesn't Fit | β | SD | HDI_3% | HDI_97% | MCSE_MEAN | MCSE_SD | ESS_Bulk | ESS_Tail | R^ |
|---|---|---|---|---|---|---|---|---|---|
| model_type[few_shot] | -1.966 | 1.657 | -5.222 | 0.967 | 0.027 | 0.017 | 3706.0 | 4357.0 | 1.00 |
| model_type[fine_tuned] | -2.331 | 1.830 | -5.742 | 1.048 | 0.028 | 0.019 | 4122.0 | 4943.0 | 1.00 |
| model_type[rag] | -4.319 | 1.556 | -7.264 | -1.398 | 0.024 | 0.016 | 4120.0 | 5219.0 | 1.00 |
| participant_type[new] | 0.531 | 0.531 | -0.489 | 1.499 | 0.018 | 0.009 | 847.0 | 1873.0 | 1.00 |
| bfpt_person_mean | -4.610 | 1.571 | -7.651 | -1.721 | 0.037 | 0.021 | 1774.0 | 2844.0 | 1.00 |
| bfpt_within | 0.024 | 0.069 | -0.108 | 0.157 | 0.001 | 0.001 | 3887.0 | 5487.0 | 1.00 |
| model_type:bfpt_person_mean[few_shot] | 4.372 | 3.445 | -1.915 | 10.898 | 0.056 | 0.036 | 3720.0 | 4061.0 | 1.00 |
| model_type:bfpt_person_mean[fine_tuned] | 6.912 | 3.779 | -0.168 | 13.898 | 0.059 | 0.040 | 4113.0 | 4861.0 | 1.00 |
| model_type:bfpt_person_mean[rag] | 9.084 | 3.222 | 3.242 | 15.395 | 0.050 | 0.033 | 4127.0 | 5247.0 | 1.00 |
| model_type:bfpt_within[few_shot] | -0.006 | 0.108 | -0.205 | 0.200 | 0.001 | 0.001 | 5319.0 | 5776.0 | 1.00 |
| model_type:bfpt_within[fine_tuned] | 0.044 | 0.129 | -0.199 | 0.290 | 0.002 | 0.001 | 5747.0 | 5423.0 | 1.00 |
| model_type:bfpt_within[rag] | -0.057 | 0.102 | -0.243 | 0.142 | 0.001 | 0.001 | 5159.0 | 5773.0 | 1.00 |

| Appropriateness Sent | β | SD | HDI_3% | HDI_97% | MCSE_MEAN | MCSE_SD | ESS_Bulk | ESS_Tail | R^ |
|---|---|---|---|---|---|---|---|---|---|
| model_type[few_shot] | 1.239 | 1.519 | -1.486 | 4.219 | 0.023 | 0.016 | 4243.0 | 4790.0 | 1.00 |
| model_type[fine_tuned] | -1.510 | 1.737 | -4.852 | 1.711 | 0.029 | 0.018 | 3533.0 | 4998.0 | 1.00 |
| model_type[rag] | 2.533 | 1.384 | -0.080 | 5.099 | 0.023 | 0.014 | 3683.0 | 4814.0 | 1.00 |
| participant_type[new] | -0.595 | 0.477 | -1.474 | 0.328 | 0.016 | 0.008 | 916.0 | 1819.0 | 1.00 |
| bfpt_person_mean | 4.546 | 1.481 | 1.876 | 7.423 | 0.037 | 0.019 | 1581.0 | 3049.0 | 1.00 |
| bfpt_within | 0.027 | 0.064 | -0.093 | 0.145 | 0.001 | 0.001 | 3972.0 | 5532.0 | 1.00 |
| model_type:bfpt_person_mean[few_shot] | -2.851 | 3.136 | -8.987 | 2.774 | 0.048 | 0.033 | 4246.0 | 4876.0 | 1.00 |
| model_type:bfpt_person_mean[fine_tuned] | 0.704 | 3.569 | -6.103 | 7.364 | 0.060 | 0.037 | 3518.0 | 4899.0 | 1.00 |
| model_type:bfpt_person_mean[rag] | -5.305 | 2.860 | -10.855 | -0.154 | 0.047 | 0.030 | 3707.0 | 4767.0 | 1.00 |
| model_type:bfpt_within[few_shot] | -0.121 | 0.098 | -0.295 | 0.072 | 0.001 | 0.001 | 4928.0 | 6183.0 | 1.00 |
| model_type:bfpt_within[fine_tuned] | -0.181 | 0.126 | -0.410 | 0.061 | 0.002 | 0.001 | 6040.0 | 6282.0 | 1.00 |
| model_type:bfpt_within[rag] | 0.039 | 0.096 | -0.141 | 0.222 | 0.001 | 0.001 | 5020.0 | 5317.0 | 1.00 |

| Appropriateness Content | β | SD | HDI_3% | HDI_97% | MCSE_MEAN | MCSE_SD | ESS_Bulk | ESS_Tail | R^ |
|---|---|---|---|---|---|---|---|---|---|
| model_type[few_shot] | 1.678 | 1.542 | -1.334 | 4.432 | 0.026 | 0.017 | 3580.0 | 4642.0 | 1.00 |
| model_type[fine_tuned] | -0.655 | 1.732 | -3.947 | 2.545 | 0.028 | 0.017 | 3941.0 | 5094.0 | 1.00 |

| | β | SD | HDI_3% | HDI_97% | MCSE_MEAN | MCSE_SD | ESS_Bulk | ESS_Tail | R^ |
|---|---|---|---|---|---|---|---|---|---|
| model_type[rag] | 2.917 | 1.432 | 0.171 | 5.549 | 0.024 | 0.016 | 3528.0 | 5151.0 | 1.00 |
| participant_type[new] | -0.473 | 0.483 | -1.368 | 0.444 | 0.016 | 0.008 | 946.0 | 1950.0 | 1.01 |
| bfpt_person_mean | 4.778 | 1.513 | 2.025 | 7.719 | 0.038 | 0.020 | 1615.0 | 3201.0 | 1.00 |
| bfpt_within | 0.030 | 0.064 | -0.091 | 0.151 | 0.001 | 0.001 | 4449.0 | 5893.0 | 1.00 |
| model_type:bfpt_person_mean[few_shot] | -3.840 | 3.187 | -9.625 | 2.233 | 0.053 | 0.035 | 3572.0 | 4734.0 | 1.00 |
| model_type:bfpt_person_mean[fine_tuned] | -1.153 | 3.561 | -7.755 | 5.561 | 0.057 | 0.036 | 3945.0 | 4825.0 | 1.00 |
| model_type:bfpt_person_mean[rag] | -6.051 | 2.957 | -11.438 | -0.352 | 0.050 | 0.033 | 3530.0 | 5146.0 | 1.00 |
| model_type:bfpt_within[few_shot] | -0.119 | 0.098 | -0.308 | 0.061 | 0.001 | 0.001 | 5564.0 | 5982.0 | 1.00 |
| model_type:bfpt_within[fine_tuned] | -0.113 | 0.122 | -0.337 | 0.120 | 0.001 | 0.001 | 6738.0 | 6228.0 | 1.00 |
| model_type:bfpt_within[rag] | 0.032 | 0.097 | -0.144 | 0.216 | 0.001 | 0.001 | 5961.0 | 6457.0 | 1.00 |

| Engaging | β | SD | HDI_3% | HDI_97% | MCSE_MEAN | MCSE_SD | ESS_Bulk | ESS_Tail | R^ |
|---|---|---|---|---|---|---|---|---|---|
| model_type[few_shot] | -0.333 | 1.501 | -3.202 | 2.413 | 0.024 | 0.015 | 4024.0 | 5366.0 | 1.00 |
| model_type[fine_tuned] | -3.086 | 1.728 | -6.308 | 0.213 | 0.027 | 0.018 | 4038.0 | 4974.0 | 1.00 |
| model_type[rag] | 1.152 | 1.357 | -1.344 | 3.666 | 0.022 | 0.014 | 3871.0 | 5265.0 | 1.00 |
| participant_type[new] | -0.310 | 0.503 | -1.241 | 0.657 | 0.018 | 0.008 | 771.0 | 2034.0 | 1.00 |

| | β | SD | HDI_3% | HDI_97% | MCSE_MEAN | MCSE_SD | ESS_Bulk | ESS_Tail | R^ |
|---|---|---|---|---|---|---|---|---|---|
| bfpt_person_mean | 2.988 | 1.513 | 0.199 | 5.943 | 0.039 | 0.020 | 1538.0 | 2654.0 | 1.00 |
| bfpt_within | 0.029 | 0.063 | -0.089 | 0.146 | 0.001 | 0.001 | 3984.0 | 5446.0 | 1.00 |
| model_type:bfpt_person_mean[few_shot] | 0.174 | 3.105 | -5.581 | 6.055 | 0.049 | 0.032 | 4052.0 | 5563.0 | 1.00 |
| model_type:bfpt_person_mean[fine_tuned] | 3.378 | 3.546 | -3.279 | 10.097 | 0.056 | 0.037 | 4021.0 | 5107.0 | 1.00 |
| model_type:bfpt_person_mean[rag] | -2.401 | 2.803 | -7.489 | 2.844 | 0.045 | 0.030 | 3897.0 | 5464.0 | 1.00 |
| model_type:bfpt_within[few_shot] | -0.092 | 0.098 | -0.275 | 0.090 | 0.001 | 0.001 | 5561.0 | 5884.0 | 1.00 |
| model_type:bfpt_within[fine_tuned] | -0.089 | 0.122 | -0.320 | 0.140 | 0.002 | 0.001 | 6355.0 | 6146.0 | 1.00 |
| model_type:bfpt_within[rag] | 0.035 | 0.094 | -0.134 | 0.221 | 0.001 | 0.001 | 4735.0 | 6080.0 | 1.00 |

| Effective | β | SD | HDI_3% | HDI_97% | MCSE_MEAN | MCSE_SD | ESS_Bulk | ESS_Tail | R^ |
|---|---|---|---|---|---|---|---|---|---|
| model_type[few_shot] | 0.195 | 1.503 | -2.513 | 3.121 | 0.024 | 0.015 | 3862.0 | 5243.0 | 1.00 |
| model_type[fine_tuned] | -2.507 | 1.737 | -5.746 | 0.800 | 0.027 | 0.018 | 4034.0 | 5198.0 | 1.00 |
| model_type[rag] | 0.751 | 1.380 | -1.768 | 3.403 | 0.022 | 0.014 | 3826.0 | 5108.0 | 1.00 |
| participant_type[new] | -0.535 | 0.553 | -1.616 | 0.480 | 0.018 | 0.010 | 921.0 | 1760.0 | 1.00 |
| bfpt_person_mean | 3.800 | 1.510 | 0.964 | 6.645 | 0.037 | 0.019 | 1636.0 | 3195.0 | 1.00 |
| bfpt_within | -0.024 | 0.064 | -0.143 | 0.094 | 0.001 | 0.001 | 3763.0 | 5449.0 | 1.00 |
| model_type:bfpt_person_mean[few_shot] | -0.858 | 3.109 | -6.787 | 4.874 | 0.050 | 0.031 | 3855.0 | 5225.0 | 1.00 |

| | β | SD | HDI_3% | HDI_97% | MCSE_MEAN | MCSE_SD | ESS_Bulk | ESS_Tail | R^ |
|---|---|---|---|---|---|---|---|---|---|
| model_type:bfpt_person_mean[fine_tuned] | 2.535 | 3.577 | -4.232 | 9.166 | 0.056 | 0.037 | 4034.0 | 5378.0 | 1.00 |
| model_type:bfpt_person_mean[rag] | -1.624 | 2.855 | -7.154 | 3.642 | 0.046 | 0.029 | 3841.0 | 5219.0 | 1.00 |
| model_type:bfpt_within[few_shot] | -0.045 | 0.098 | -0.237 | 0.127 | 0.001 | 0.001 | 4726.0 | 5578.0 | 1.00 |
| model_type:bfpt_within[fine_tuned] | -0.058 | 0.126 | -0.283 | 0.186 | 0.002 | 0.001 | 6115.0 | 6127.0 | 1.00 |
| model_type:bfpt_within[rag] | 0.075 | 0.096 | -0.107 | 0.253 | 0.001 | 0.001 | 5035.0 | 5864.0 | 1.00 |

| Professional | β | SD | HDI_3% | HDI_97% | MCSE_MEAN | MCSE_SD | ESS_Bulk | ESS_Tail | R^ |
|---|---|---|---|---|---|---|---|---|---|
| model_type[few_shot] | 0.647 | 1.531 | -2.247 | 3.478 | 0.025 | 0.017 | 3775.0 | 5387.0 | 1.00 |
| model_type[fine_tuned] | -2.685 | 1.736 | -5.767 | 0.653 | 0.028 | 0.019 | 3897.0 | 4524.0 | 1.00 |
| model_type[rag] | 2.762 | 1.422 | 0.107 | 5.466 | 0.024 | 0.014 | 3518.0 | 5312.0 | 1.00 |
| participant_type[new] | -0.709 | 0.513 | -1.633 | 0.288 | 0.018 | 0.009 | 837.0 | 1640.0 | 1.01 |
| bfpt_person_mean | 5.336 | 1.551 | 2.424 | 8.177 | 0.045 | 0.021 | 1203.0 | 2265.0 | 1.00 |
| bfpt_within | -0.016 | 0.065 | -0.140 | 0.102 | 0.001 | 0.001 | 3518.0 | 5155.0 | 1.00 |
| model_type:bfpt_person_mean[few_shot] | -1.645 | 3.165 | -7.661 | 4.168 | 0.052 | 0.034 | 3754.0 | 5397.0 | 1.00 |
| model_type:bfpt_person_mean[fine_tuned] | 3.117 | 3.569 | -3.585 | 9.567 | 0.057 | 0.038 | 3903.0 | 4663.0 | 1.00 |
| model_type:bfpt_person_mean[rag] | -5.684 | 2.937 | -11.201 | -0.163 | 0.050 | 0.030 | 3509.0 | 5184.0 | 1.00 |

| | β | SD | HDI_3% | HDI_97% | MCSE_MEAN | MCSE_SD | ESS_Bulk | ESS_Tail | R^ |
|---|---|---|---|---|---|---|---|---|---|
| model_type:bfpt_within[few_shot] | -0.038 | 0.102 | -0.231 | 0.153 | 0.001 | 0.001 | 4936.0 | 5740.0 | 1.00 |
| model_type:bfpt_within[fine_tuned] | 0.001 | 0.123 | -0.232 | 0.224 | 0.002 | 0.001 | 5318.0 | 5689.0 | 1.00 |
| model_type:bfpt_within[rag] | -0.007 | 0.098 | -0.197 | 0.172 | 0.001 | 0.001 | 4680.0 | 5747.0 | 1.00 |

| Angry | β | SD | HDI_3% | HDI_97% | MCSE_MEAN | MCSE_SD | ESS_Bulk | ESS_Tail | R^ |
|---|---|---|---|---|---|---|---|---|---|
| model_type[few_shot] | -0.454 | 2.077 | -4.306 | 3.431 | 0.033 | 0.021 | 3941.0 | 4599.0 | 1.00 |
| model_type[fine_tuned] | 1.754 | 2.177 | -2.192 | 5.950 | 0.032 | 0.022 | 4509.0 | 5347.0 | 1.00 |
| model_type[rag] | -2.371 | 1.956 | -6.038 | 1.295 | 0.030 | 0.020 | 4371.0 | 5138.0 | 1.00 |
| participant_type[new] | 0.267 | 0.578 | -0.862 | 1.321 | 0.016 | 0.009 | 1373.0 | 2532.0 | 1.01 |
| bfpt_person_mean | -8.092 | 2.228 | -12.408 | -4.114 | 0.057 | 0.031 | 1530.0 | 2610.0 | 1.00 |
| bfpt_within | -0.060 | 0.093 | -0.229 | 0.119 | 0.001 | 0.001 | 3876.0 | 5302.0 | 1.00 |
| model_type:bfpt_person_mean[few_shot] | 1.106 | 4.281 | -6.977 | 8.952 | 0.068 | 0.044 | 3945.0 | 4553.0 | 1.00 |
| model_type:bfpt_person_mean[fine_tuned] | -1.786 | 4.482 | -10.423 | 6.306 | 0.067 | 0.046 | 4513.0 | 5064.0 | 1.00 |
| model_type:bfpt_person_mean[rag] | 4.813 | 4.034 | -2.619 | 12.476 | 0.061 | 0.041 | 4368.0 | 5215.0 | 1.00 |
| model_type:bfpt_within[few_shot] | 0.137 | 0.140 | -0.127 | 0.397 | 0.002 | 0.001 | 4745.0 | 5741.0 | 1.00 |
| model_type:bfpt_within[fine_tuned] | -0.116 | 0.158 | -0.398 | 0.193 | 0.002 | 0.002 | 5483.0 | 6230.0 | 1.00 |

| | β | SD | HDI_3% | HDI_97% | MCSE_MEAN | MCSE_SD | ESS_Bulk | ESS_Tail | R^ |
|---|---|---|---|---|---|---|---|---|---|
| model_type:bfpt_within[rag] | 0.059 | 0.138 | -0.198 | 0.322 | 0.002 | 0.001 | 5133.0 | 6404.0 | 1.00 |

| Annoyed | β | SD | HDI_3% | HDI_97% | MCSE_MEAN | MCSE_SD | ESS_Bulk | ESS_Tail | R^ |
|---|---|---|---|---|---|---|---|---|---|
| model_type[few_shot] | -1.488 | 1.852 | -5.100 | 1.895 | 0.032 | 0.019 | 3294.0 | 4706.0 | 1.00 |
| model_type[fine_tuned] | 3.020 | 1.961 | -0.703 | 6.659 | 0.034 | 0.020 | 3383.0 | 4953.0 | 1.00 |
| model_type[rag] | -1.455 | 1.752 | -4.748 | 1.755 | 0.030 | 0.018 | 3526.0 | 4713.0 | 1.00 |
| participant_type[new] | -0.278 | 0.513 | -1.234 | 0.666 | 0.016 | 0.008 | 1087.0 | 2024.0 | 1.00 |
| bfpt_person_mean | -4.154 | 2.139 | -8.250 | -0.241 | 0.055 | 0.032 | 1505.0 | 2335.0 | 1.00 |
| bfpt_within | -0.098 | 0.081 | -0.254 | 0.049 | 0.001 | 0.001 | 3854.0 | 5072.0 | 1.00 |
| model_type:bfpt_person_mean[few_shot] | 3.429 | 3.821 | -3.831 | 10.584 | 0.066 | 0.039 | 3316.0 | 4678.0 1.00 | |
| model_type:bfpt_person_mean[fine_tuned] | -4.357 | 4.033 | -11.945 | 3.236 | 0.069 | 0.041 | 3387.0 | 5086.0 | 1.00 |
| model_type:bfpt_person_mean[rag] | 3.039 | 3.615 | -3.674 | 9.818 | 0.061 | 0.038 | 3536.0 | 4877.0 | 1.00 |
| model_type:bfpt_within[few_shot] | 0.101 | 0.124 | -0.131 | 0.332 | 0.002 | 0.001 | 5261.0 | 5775.0 | 1.00 |
| model_type:bfpt_within[fine_tuned] | -0.004 | 0.139 | -0.266 | 0.257 | 0.002 | 0.001 | 5179.0 | 5913.0 | 1.00 |
| model_type:bfpt_within[rag] | 0.048 | 0.119 | -0.166 | 0.281 | 0.002 | 0.001 | 5110.0 | 5932.0 | 1.00 |

| Frustrated | β | SD | HDI_3% | HDI_97% | MCSE_MEAN | MCSE_SD | ESS_Bulk | ESS_Tail | R^ |
|---|---|---|---|---|---|---|---|---|---|

| | β | SD | HDI_3% | HDI_97% | MCSE_MEAN | MCSE_SD | ESS_Bulk | ESS_Tail | R^ |
|---|---|---|---|---|---|---|---|---|---|
| model_type[few_shot] | -0.892 | 1.971 | -4.507 | 2.855 | 0.033 | 0.021 | 3584.0 | 5130.0 | 1.0 |
| model_type[fine_tuned] | -0.158 | 2.056 | -3.897 | 3.822 | 0.034 | 0.021 | 3715.0 | 5334.0 | 1.0 |
| model_type[rag] | -2.122 | 1.845 | -5.697 | 1.174 | 0.031 | 0.019 | 3636.0 | 5033.0 | 1.0 |
| participant_type[new] | -0.004 | 0.488 | -0.905 | 0.917 | 0.014 | 0.008 | 1219.0 | 2455.0 | 1.0 |
| bfpt_person_mean | -6.653 | 1.865 | -10.152 | -3.148 | 0.046 | 0.024 | 1629.0 | 3025.0 | 1.0 |
| bfpt_within | 0.064 | 0.081 | -0.081 | 0.221 | 0.001 | 0.001 | 3328.0 | 5057.0 | 1.0 |
| model_type:bfpt_person_mean[few_shot] | 1.745 | 4.064 | -6.060 | 9.061 | 0.068 | 0.043 | 3565.0 | 4840.0 | 1.0 |
| model_type:bfpt_person_mean[fine_tuned] | 2.056 | 4.234 | -5.981 | 9.889 | 0.069 | 0.043 | 3718.0 | 5393.0 | 1.0 |
| model_type:bfpt_person_mean[rag] | 4.464 | 3.812 | -2.356 | 11.818 | 0.063 | 0.038 | 3631.0 | 4902.0 | 1.0 |
| model_type:bfpt_within[few_shot] | 0.023 | 0.124 | -0.198 | 0.271 | 0.002 | 0.001 | 4602.0 | 5601.0 | 1.0 |
| model_type:bfpt_within[fine_tuned] | -0.157 | 0.139 | -0.414 | 0.104 | 0.002 | 0.001 | 5327.0 | 5749.0 | 1.0 |
| model_type:bfpt_within[rag] | -0.108 | 0.121 | -0.337 | 0.116 | 0.002 | 0.001 | 4283.0 | 5409.0 | 1.0 |

| Happy | β | SD | HDI_3% | HDI_97% | MCSE_MEAN | MCSE_SD | ESS_Bulk | ESS_Tail | R^ |
|---|---|---|---|---|---|---|---|---|---|
| model_type[few_shot] | 0.863 | 1.501 | -1.817 | 3.792 | 0.023 | 0.015 | 4339.0 | 5245.0 | 1.00 |
| model_type[fine_tuned] | -2.301 | 1.729 | -5.413 | 1.064 | 0.028 | 0.018 | 3886.0 | 4979.0 | 1.00 |

| | β | SD | HDI_3% | HDI_97% | MCSE_MEAN | MCSE_SD | ESS_Bulk | ESS_Tail | R^ |
|---|---|---|---|---|---|---|---|---|---|
| model_type[rag] | 1.936 | 1.390 | -0.747 | 4.494 | 0.021 | 0.014 | 4422.0 | 5129.0 | 1.00 |
| participant_type[new] | 0.067 | 0.483 | -0.857 | 0.978 | 0.016 | 0.009 | 909.0 | 1923.0 | 1.00 |
| bfpt_person_mean | 2.899 | 1.462 | 0.188 | 5.628 | 0.038 | 0.018 | 1516.0 | 3143.0 | 1.00 |
| bfpt_within | -0.015 | 0.064 | -0.135 | 0.106 | 0.001 | 0.001 | 3934.0 | 5396.0 | 1.00 |
| model_type:bfpt_person_mean[few_shot] | -1.914 | 3.101 | -7.938 | 3.682 | 0.047 | 0.032 | 4338.0 | 5350.0 | 1.00 |
| model_type:bfpt_person_mean[fine_tuned] | 2.922 | 3.545 | -3.916 | 9.346 | 0.057 | 0.036 | 3891.0 | 4996.0 | 1.00 |
| model_type:bfpt_person_mean[rag] | -3.852 | 2.867 | -9.019 | 1.835 | 0.043 | 0.030 | 4418.0 | 5093.0 | 1.00 |
| model_type:bfpt_within[few_shot] | -0.036 | 0.099 | -0.223 | 0.144 | 0.001 | 0.001 | 5249.0 | 5949.0 | 1.00 |
| model_type:bfpt_within[fine_tuned] | -0.061 | 0.122 | -0.294 | 0.167 | 0.002 | 0.001 | 5798.0 | 6110.0 | 1.00 |
| model_type:bfpt_within[rag] | 0.059 | 0.095 | -0.128 | 0.229 | 0.001 | 0.001 | 5101.0 | 5783.0 | 1.00 |

| Sad | β | SD | HDI_3% | HDI_97% | MCSE_MEAN | MCSE_SD | ESS_Bulk | ESS_Tail | R^ |
|---|---|---|---|---|---|---|---|---|---|
| Model_type[few_shot] | -1.538 | 2.196 | -5.610 | 2.510 | 0.037 | 0.024 | 3559.0 | 4740.0 | 1.00 |
| model_type[fine_tuned] | 0.399 | 2.386 | -4.119 | 4.911 | 0.038 | 0.026 | 4042.0 | 4925.0 | 1.00 |
| model_type[rag] | -2.412 | 2.068 | -6.265 | 1.483 | 0.034 | 0.022 | 3604.0 | 5052.0 | 1.00 |
| participant_type[new] | -0.345 | 0.685 | -1.620 | 0.924 | 0.019 | 0.010 | 1314.0 | 2206.0 | 1.00 |

| | β | SD | HDI_3% | HDI_97% | MCSE_MEAN | MCSE_SD | ESS_Bulk | ESS_Tail | R^ |
|---|---|---|---|---|---|---|---|---|---|
| bfpt_person_mean | -10.789 | 2.674 | -16.024 | -6.014 | 0.069 | 0.039 | 1515.0 | 2382.0 | 1.00 |
| bfpt_within | 0.083 | 0.104 | -0.113 | 0.279 | 0.002 | 0.001 | 3518.0 | 5099.0 | 1.00 |
| model_type:bfpt_person_mean[few_shot] | 3.533 | 4.522 | -4.927 | 11.800 | 0.077 | 0.049 | 3506.0 | 4844.0 | 1.00 |
| model_type:bfpt_person_mean[fine_tuned] | 0.352 | 4.885 | -8.708 | 9.845 | 0.077 | 0.054 | 4035.0 | 4638.0 | 1.00 |
| model_type:bfpt_person_mean[rag] | 4.996 | 4.252 | -2.932 | 12.989 | 0.071 | 0.044 | 3574.0 | 5316.0 | 1.00 |
| model_type:bfpt_within[few_shot] | -0.079 | 0.155 | -0.370 | 0.207 | 0.002 | 0.001 | 4406.0 | 5613.0 | 1.00 |
| model_type:bfpt_within[fine_tuned] | -0.149 | 0.179 | -0.489 | 0.181 | 0.002 | 0.002 | 5155.0 | 5670.0 | 1.00 |
| model_type:bfpt_within[rag] | 0.095 | 0.151 | -0.185 | 0.382 | 0.002 | 0.001 | 4843.0 | 5849.0 | 1.00 |

| Scared | β | SD | HDI_3% | HDI_97% | MCSE_MEAN | MCSE_SD | ESS_Bulk | ESS_Tail | R^ |
|---|---|---|---|---|---|---|---|---|---|
| model_type[few_shot] | -1.126 | 2.366 | -5.353 | 3.493 | 0.033 | 0.026 | 5010.0 | 5335.0 | 1.0 |
| model_type[fine_tuned] | 0.870 | 2.590 | -4.053 | 5.609 | 0.036 | 0.029 | 5085.0 | 5192.0 | 1.0 |
| model_type[rag] | -1.626 | 2.182 | -5.625 | 2.544 | 0.033 | 0.025 | 4517.0 | 4662.0 | 1.0 |
| participant_type[new] | 0.209 | 1.067 | -1.726 | 2.325 | 0.025 | 0.016 | 1888.0 | 3298.0 | 1.0 |
| bfpt_person_mean | -16.265 | 4.965 | -25.683 | -7.018 | 0.110 | 0.078 | 2064.0 | 2707.0 | 1.0 |
| bfpt_within | 0.028 | 0.122 | -0.197 | 0.258 | 0.002 | 0.001 | 4522.0 | 5163.0 | 1.0 |
| model_type:bfpt_person_mean[few_shot] | 2.875 | 4.854 | -6.552 | 11.642 | 0.068 | 0.053 | 5043.0 | 5281.0 | 1.0 |

| | β | SD | HDI_3% | HDI_97% | MCSE_MEAN | MCSE_SD | ESS_Bulk | ESS_Tail | R^ |
|---|---|---|---|---|---|---|---|---|---|
| model_type:bfpt_person_mean[fine_tuned] | -0.669 | 5.279 | -10.824 | 8.854 | 0.074 | 0.058 | 5110.0 | 5226.0 | 1.0 |
| model_type:bfpt_person_mean[rag] | 3.866 | 4.484 | -4.614 | 12.127 | 0.067 | 0.050 | 4537.0 | 5016.0 | 1.0 |
| model_type:bfpt_within[few_shot] | -0.101 | 0.176 | -0.436 | 0.219 | 0.002 | 0.002 | 5671.0 | 5822.0 | 1.0 |
| model_type:bfpt_within[fine_tuned] | -0.246 | 0.207 | -0.638 | 0.133 | 0.003 | 0.002 | 6480.0 | 6411.0 | 1.0 |
| model_type:bfpt_within[rag] | 0.106 | 0.174 | -0.229 | 0.421 | 0.002 | 0.002 | 5826.0 | 5383.0 | 1.0 |

| Surprised | β | SD | HDI_3% | HDI_97% | MCSE_MEAN | MCSE_SD | ESS_Bulk | ESS_Tail | R^ |
|---|---|---|---|---|---|---|---|---|---|
| model_type[few_shot] | -0.154 | 1.670 | -3.319 | 2.940 | 0.025 | 0.017 | 4459.0 | 5290.0 | 1.00 |
| model_type[fine_tuned] | 3.300 | 1.927 | -0.475 | 6.714 | 0.029 | 0.020 | 4338.0 | 4935.0 | 1.00 |
| model_type[rag] | 0.257 | 1.531 | -2.636 | 3.106 | 0.023 | 0.015 | 4503.0 | 5535.0 | 1.00 |
| participant_type[new] | 0.975 | 0.568 | -0.104 | 2.072 | 0.020 | 0.010 | 789.0 | 1668.0 | 1.01 |
| bfpt_person_mean | -2.031 | 1.554 | -4.994 | 0.888 | 0.040 | 0.019 | 1498.0 | 3090.0 | 1.00 |
| bfpt_within | -0.001 | 0.068 | -0.128 | 0.125 | 0.001 | 0.001 | 4156.0 | 5404.0 | 1.00 |
| model_type:bfpt_person_mean[few_shot] | 0.097 | 3.431 | -6.469 | 6.410 | 0.052 | 0.035 | 4435.0 | 5328.0 | 1.00 |
| model_type:bfpt_person_mean[fine_tuned] | -7.366 | 3.946 | -14.812 | -0.082 | 0.060 | 0.040 | 4349.0 | 4895.0 | 1.00 |
| model_type:bfpt_person_mean[rag] | -0.686 | 3.151 | -6.705 | 5.146 | 0.047 | 0.031 | 4490.0 | 5424.0 | 1.00 |

| | | | | | | | | |
|---|---|---|---|---|---|---|---|---|
| model_type:bfpt_within[few_shot] | -0.038 | 0.108 | -0.241 | 0.165 | 0.001 | 0.001 | 5975.0 | 5913.0 | 1.00 |
| model_type:bfpt_within[fine_tuned] | -0.074 | 0.131 | -0.326 | 0.166 | 0.002 | 0.001 | 7042.0 | 5970.0 | 1.00 |
| model_type:bfpt_within[rag] | -0.074 | 0.103 | -0.268 | 0.118 | 0.001 | 0.001 | 5539.0 | 5812.0 | 1.00 |